\documentclass[superscriptaddress,twocolumn,showpacs]{revtex4}
\usepackage{amsmath}
\usepackage{amssymb}
\usepackage{bm}
\usepackage{epsfig}
\usepackage{graphicx}
\usepackage [russian,english]{babel}
\usepackage{color}

\renewcommand{\d}{\mathrm d}

\begin{document}

\newcount\timehh  \newcount\timemm
\timehh=\time \divide\timehh by 60
\timemm=\time
\count255=\timehh\multiply\count255 by -60 \advance\timemm by \count255

\title{Theory of optical spin control in quantum dot microcavities}

\author{D.~S.~Smirnov}
\affiliation{Ioffe Institute of the RAS, 194021, St.-Petersburg, Russia}

\author{M.~M.~Glazov}
\affiliation{Ioffe Institute of the RAS, 194021, St.-Petersburg, Russia}

\author{E.~L.~Ivchenko}
\affiliation{Ioffe Institute of the RAS, 194021, St.-Petersburg, Russia}

\author{L.~Lanco}
\affiliation{Laboratoire de Photonique et de Nanostructures, CNRS UPR 20, Route de Nozay, 91460 Marcoussis, France}
\affiliation{Universit\'e Paris Diderot --- Paris 7, 75205 Paris CEDEX 13, France}


\begin{abstract}
We present a microscopic theory of optical initialization, control and detection for a single electron spin in a quantum dot embedded into a zero-dimensional microcavity. The strong coupling regime of the trion and the cavity mode is addressed. We demonstrate that efficient spin orientation by a single circularly polarized pulse is possible in relatively weak transverse magnetic fields. The possibilities for spin control by additional circularly polarized pulse are analyzed. Under optimal conditions the Kerr and Faraday rotation angles induced by the spin polarized electron may reach tens of degrees.
\end{abstract}

\maketitle


\section{Introduction}\label{sec:intro}

Non-magnetic spin control is among the most rapidly developing topics of modern semiconductor spin physics. Substantial progress has been achieved by application of the pump-probe technique to bulk semiconductors, quantum wells and quantum dots, where spins are created, manipulated and detected by optical pulses~\cite{dyakonov_book}. Singly charged quantum dot structures are of particular interest. In these systems discrete energy spectrum of charge carriers makes resonant optical excitation possible, which substantially enhances spin-photon coupling. At the same time, quenching the orbital motion due to size quantization suppresses significantly the spin-orbit induced spin decoherence making it possible to achieve long spin lifetimes of single electrons or holes. In addition to robust optical initialization of electron spin by polarized light, spin polarization readout~\cite{PhysRevLett.94.047402,greilich06}, and ultrafast spin rotation by light~\cite{Greilich2009}, in quantum dot ensembles a number of prominent effects, such as spin precession mode-locking and nuclei-induced frequency focussing have been  demonstrated~\cite{A.Greilich07212006,A.Greilich09282007}, see, e.g.,~Ref.~\cite{glazov:review} for review.

The spin-photon interfacing and optical spin control has also been demonstrated for single quantum dots~\cite{MeteAtature04282006,J.Berezovsky04182008,mikkelsen07,atature07,2011NaPho...5..702G}. The charge carrier spin state is, as a rule, detected by the Faraday or Kerr rotation. Typically, the Faraday rotation angle was on the order of $10^{-3}$ degrees in Refs.~\cite{J.Berezovsky04182008,mikkelsen07,atature07}. The efficiency of spin-photon interaction can be strongly enhanced by placing the active system into a microcavity, where the electric field and, accordingly, Faraday/Kerr rotation are accumulated due to multiple passages of light between the mirrors~\cite{Kavokin:1997fk,PhysRevB.78.085307}. It makes possible to reach giant spin-Faraday effect in bulk semiconductor~\cite{PhysRevB.85.195313}, and even to detect a host-lattice nuclei-induced Faraday rotation~\cite{PhysRevLett.111.087603}. On a single spin level embedding a quantum dot QD into a microcavity has enabled to achieve a resident hole spin control~\cite{De-Greve:2011uq} and detect fluctuations of a single spin~\cite{PhysRevLett.112.156601}. Very recently, a macroscopic rotation of photon polarization reaching several degrees has been observed in a single QD deterministically coupled to a photonic mode of a micropillar cavity demonstrating efficient spin-photon coupling~\cite{Arnold2015}. 

In quantum-dot/microcavity structures two regimes of light-matter interaction, namely, weak and strong coupling, are known~\cite{andreani99a,Khitrova2006,microcavities}. In the weak coupling regime, realized in above works~\cite{De-Greve:2011uq,PhysRevLett.112.156601,Arnold2015}, the dampings of cavity mode and trion exceed their coupling constant, while in the strong coupling regime the dampings are small as compared with the coupling constant, see Ref.~\cite{ivchenko05a} for rigorous criterion, and the coherent energy transfer between the photon and material excitation becomes possible. The strong coupling regime has been already achieved for neutral QDs placed in various types of microcavities~\cite{yoshie04a,reithmaier04a,peter05a,hennessy07}, see Ref.~\cite{Khitrova2006} for review. 

The strong coupling between a two-level quantum system and a single electromagnetic mode in a zero-dimensional cavity results in a formation of the Jaynes√Cummings ladder 
of coupled states~\cite{JC}. The rung splitting in this ladder is dependent on the number of photon quanta in the cavity. The spectroscopic manifestation of this dependence carries a direct signature of the quantization of light and reveals non-semiclassical properties of the intracavity field as observed in atomic systems~\cite{atomcavity}, superconducting circuits~\cite{fink}, and in microcavities with neutral quantum dots, where zero-dimensional exciton provides the two-level atom-like nonlinearity~\cite{dotcavity}.

Fundamental limitations on realization of the strong coupling regime for the charged QDs embedded into microcavities are absent~\cite{Arnold2015}: The light-matter interaction is accompanied by generation of trions, three particle Fermionic complexes with two electrons in the spin singlet state and a hole with an unpaired spin in case of the resident electron ($X^-$ trion) or with two holes and unpaired electron in case of the resident hole ($X^+$ trion). This brings spin degree of freedom into the Jaynes-Cummings ladder and calls for a theory of the optical control of a single spin in the QD cavity quantum-electrodynamics structure operating in the strong coupling regime. Such a theory is developed in the present paper. We demonstrate a possibility to initialize the electron spin by a single circularly polarized pulse in the transverse magnetic field. It is also shown that the train of pulses leads to a complete spin orientation. Moreover, we predict an efficient spin rotation by the circularly polarized control pulse and tens of degrees for the spin-Kerr and Faraday rotation angles of the probe pulse.

\section{Model}\label{sec:model}

\begin{figure}[t]
\includegraphics[width=\linewidth]{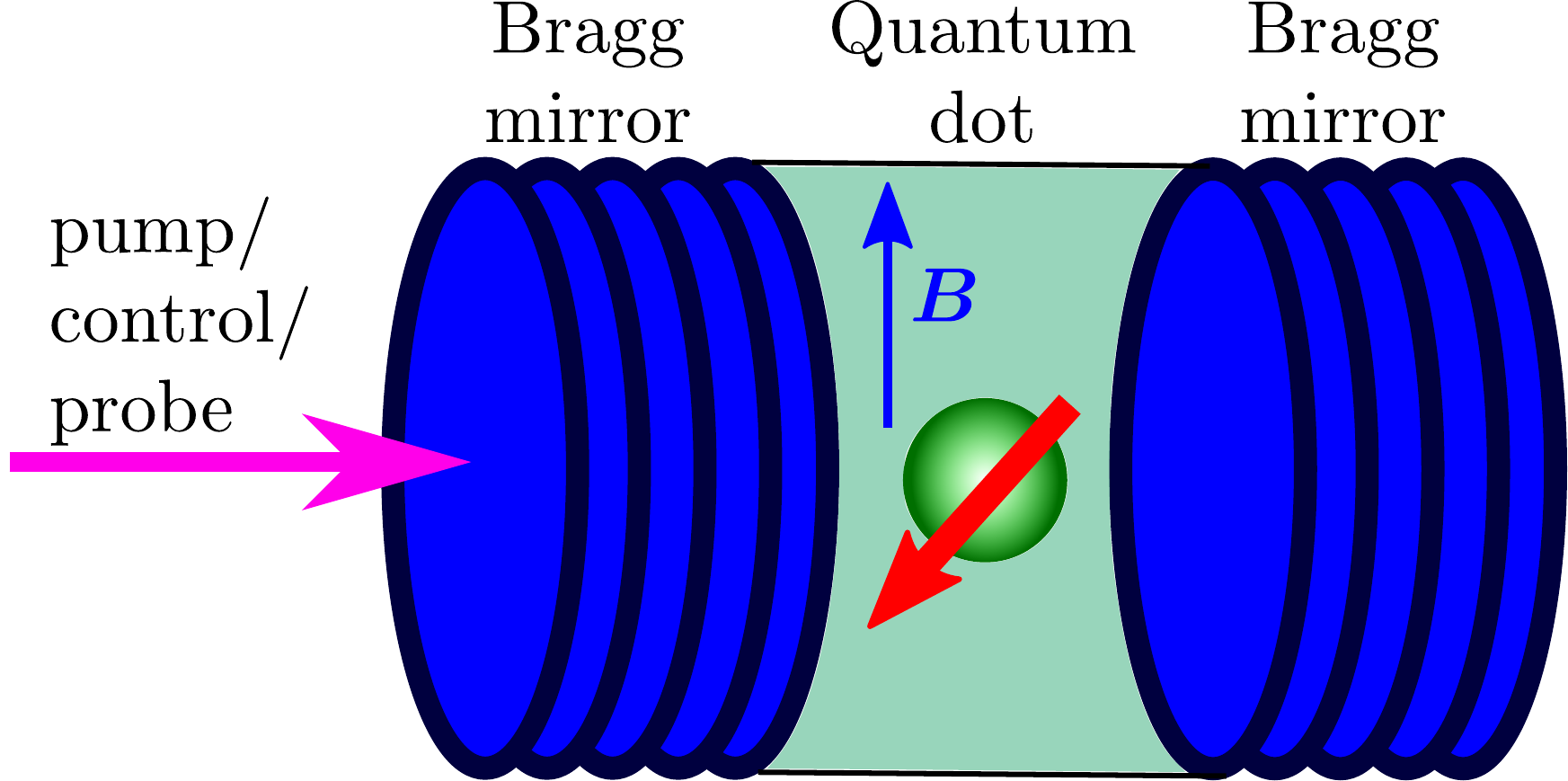}
\caption{(a) Sketch of the single quantum dot microcavity: green sphere depicts the quantum dot, red arrow depicts the single localized electron spin.}
\label{fig:scheme:sys}
\end{figure}

We consider a zero-dimensional microcavity with an embedded negatively charged QD. We denote the growth axis as $z$ axis, and assume that the light is incident along~$z$. The scheme of the system under study is sketched in Fig.~\ref{fig:scheme:sys}. The cavity eigenfrequency $\omega_c$ is assumed to be close to the trion resonant frequency $\omega_0$ in the dot, and for simplicity we assume the two orthogonally polarized cavity modes to be degenerate. In the presence of an external magnetic field $\bm B$ applied along the $x$-axis, the Hamiltonian of the system 
\begin{equation}
 \mathcal H=\mathcal H_++\mathcal H_-+\mathcal H_B
 \label{ham}
\end{equation} 
is written as a sum of the light-matter interaction operator 
\begin{multline}
 \mathcal{H}_\pm = \hbar\omega_c c_\pm^\dag c_\pm + \hbar\omega_0 a_{\pm \frac32}^\dag a_{\pm \frac32} 
 \\  + \left( \hbar gc_i^\dag a_{\pm \frac12}^\dag a_{\pm \frac32} + \hbar\mathcal{E}_\pm(t) e^{-i\omega t} c_\pm^\dag + {\rm h.c.} \right)
 \label{hpm}
\end{multline}
and the Zeeman Hamiltonian
\begin{equation}
\mathcal{H}_B=\frac{\hbar\Omega}{2}(a^\dag_{\frac12}a_{- \frac12}+a^\dag_{- \frac12}a_{\frac12})\:.
\end{equation}
Here $c_\pm^\dag, c_\pm$ are the creation and annihilation operators for the $\sigma^\pm$ circularly-polarized cavity photons, $a_{S_z}^\dag, a_{S_z}$ and $a_{J_z}^\dag, a_{J_z}$ are the creation and annihilation operators, respectively, for the electron with the spin $z$-component $S_z = \pm 1/2$ and the singlet trion with the heavy-hole angular-momentum projection $J_z = \pm 3/2$, $\hbar g$ is the photon-trion coupling constant, $\Omega = g_e \mu_B B/\hbar$ is the Larmor frequency of the electron spin precession in the magnetic field, $g_e$ is the electron $g$-factor describing the Zeeman effect.  Due to a small value of the transverse L\'ande factor for heavy holes \cite{Mar99} we neglect, hereafter, the trion spin precession. We assume the QD size to be small compared with the exciton Bohr radius which allows us to treat the trions (three-particle complexes) as Fermions~\cite{Poddubny2010} and use for $a_{J_z}^\dag, a_{J_z}$ the Fermionic commutation relations. Furthermore, the Hamiltonian~(\ref{hpm}) includes an external electric field of the pump, probe or control pulse with the carrier frequency $\omega$ and smooth real envelopes $\mathcal E_\pm(t)$ being proportional to the electric field incident on the zero-dimensional microcavity and mirror transmission coefficient~\cite{milburn}.

Note that although the incident fields $\mathcal E_\pm(t)$ are treated classically we take into account the quantization of photonic states inside the cavity. The Hilbert space corresponding to this model is thus constituted by the four-level spin system of the QD and by the Fock states of the intracavity $\sigma^+$ and $\sigma^-$ photons. Regarding the spin-system, its four basis states are denoted $|\uparrow\rangle$, $|\downarrow\rangle$ (corresponding to an electron with $S_z = 1/2$ or $S_z=-1/2$, respecively) and $|\Uparrow\rangle$, $|\Downarrow\rangle$ (corresponding to a trion with $J_z = 3/2$ or $J_z=-3/2$), and the basis states for the cavity are denoted $|m_+\rangle$ and $|m_-\rangle$, with $m_{\pm}=0,1,2...$ the respective numbers of $\sigma^+$ and $\sigma^-$ photons. In the following we use notations such as $|\uparrow,m_+,m_-\rangle$ to denote a state of the QD/microcavity system with spin-up electron and photon numbers $m_+$ and $m_-$ \cite{Arnold2015}.

The Hamiltonian~\eqref{ham} describes the coherent energy exchange between photons and trions in the microcavity as well as the effects of external pumping and magnetic field. In order to describe incoherent processes of photon mode decay via the photon leakage through the mirrors and non-radiative decay of trions in QDs we should introduce the density matrix $\rho$ satisfying the quantum master equation
\begin{equation}
\label{density:m}
  \dot\rho(t)=i[\rho(t),\mathcal H]-\mathcal L\lbrace\rho(t)\rbrace\:,
\end{equation}
where $\dot\rho(t)$ denotes the time-derivative of the density matrix and the Lindblad operator $\mathcal L\lbrace\rho(t)\rbrace$ responsible for the incoherent processes can be presented as~\cite{Poddubny2010,milburn,Carmichael}
\begin{multline}
\label{Lindblad}
\mathcal L\lbrace\rho(t)\rbrace = \sum_{j=\pm} \left\{ \varkappa \left[ c_j^\dag c_j\rho(t)+\rho(t) c_j^\dag c_j-2c_j\rho(t) c_j^\dag \right] \right. \\
+\ \gamma \left[ a_{j \frac32}^\dag a_{j \frac32}\rho(t) + \rho(t) a_{j \frac32}^\dag a_{j \frac32} \right. \\
\left. \left. -\ 2\ a_{j \frac12}^\dag a_{j\frac32}\rho(t) a_{j \frac32}^\dag a_{ j \frac12} \right] \right\}\:. \hspace{1.8 cm} \mbox{} 
\end{multline}
Here $\varkappa$ is the decay rate of the cavity mode, and $\gamma$ is the trion decay rate unrelated with the photon emission to the cavity mode. 
{In this paper, we assume the former to exceed, by far, the latter ($\varkappa \gg \gamma$) as it is typically the case~\cite{deph2,hennessy07}, focus on a quite strong coupling regime ($\gamma, \varkappa \ll g$) and consider moderately high magnetic fields:
\begin{equation}
\label{b:field}
\gamma \ll \varkappa \ll \Omega \ll g\:.
\end{equation}
In this model the initialization, control and detection of the single electron spin in the cavity is practically independent of $\gamma$, and the inclusion of this additional parameter allows us to check its low influence.} 

As will be shown below, the condition $\Omega \ll g$ allows us to neglect the magnetic-field induced mixing between the excited states and take it into account only for the ground states $|\uparrow,0,0\rangle$ and $|\downarrow,0,0\rangle$. In such a regime, the Hamiltonian $\mathcal{H}_B$ leads to coherent spin beats at the frequency $\Omega$ between these two ground states, whereas all the other states remain negligibly affected~\footnote{{The magnetic field also affects the states with same number of $\sigma^+$ and $\sigma^-$ photons, i.e., $|\uparrow,m,m\rangle$ and $|\downarrow,m,m\rangle$. Under the considered conditions these states can be realized for linearly or elliptically polarized incident light, i.e., for the probe. However, since the probe is weak, this mixing can be disregarded, see Sec.~\ref{sec:det}.}}. Furthermore, if $\gamma, \varkappa \ll \Omega$ the spin precession is much faster than the trion or photon decay processes, the limit in which the spin dynamics is governed only by the population and coherence of the two ground states. For reasonable values of the parameters $\hbar g=500$~$\mu$eV, $\hbar\varkappa = 10$~$\mu$eV and $\hbar\gamma=2~\mu$eV, the magnetic field satisfying the conditions~(\ref{b:field}) corresponds to the range of 1$\div$10 T. 

{Additional pure dephasing processes, with a dephasing rate $\gamma^*$, can also be included into Eq.~\eqref{Lindblad} following, e.g., Ref.~\cite{deph2,deph1,deph3}. This increases the relaxation rate of the off-diagonal elements from the rate $\gamma/2$ to $\gamma^* + \gamma/2$. However, in the strong-coupling regime with $\gamma$ and $\gamma^*$ being much smaller than $\varkappa$ the results are almost insensitive to both $\gamma$ and $\gamma^*$.} 

{A general description of the decay and decoherence processes would require the inclusion of the relaxation and dephasing sources (i.e., electron-phonon interaction, coupling with other photon modes, etc.) in the Hamiltonian of a system~\cite{milburn,Carmichael,pulsed,non-markovian}. However, in the present study we restrict the analysis to the simplest possible model, considering only the Markovian processes that can be described by Eq.~\eqref{Lindblad}. In the following the spin relaxation processes of electron and trion (e.g., splin-flip and/or spin-decoherence experienced by the hole spin in the trion state, even in the absence of trion decay) are also neglected because for the zero-dimensional carrier states they are slow as compared with the trion decay rate $\gamma$ and the photon decay rate $\varkappa$ (see Refs.~\cite{A.Greilich07212006,glazov:review} and references therein). Accordingly, they play no role in the spin orientation and control. It is worth to mention that the formalism of Eqs.~\eqref{ham}, \eqref{density:m} and \eqref{Lindblad} is derived under the conditions $\gamma, \varkappa \ll \omega_c,\omega_0,\omega$, i.e. when the incoherent processes are perturbative, and $g \ll \omega_c,\omega_0,\omega$, while the relations between the decay rates $\gamma, \varkappa$ and the photon-trion coupling constant $g$ can be arbitrary. Hence, the formalism is applicable in both the weak and strong coupling regimes~\cite{just}.}

\section{Spin initialization and control}\label{sec:ic}

In this section we study the effect of a short circularly-polarized pulse on the resident electron spin in the quantum microcavity. 
First we present the formalism based on the Schr\"{o}dinger equation to describe the action of pump and control pulses. Next we consider the spin initialization process and then address the coherent spin control by circularly polarized pulses.

\subsection{Formalism}
We bound ourselves to short optical pulses with a duration $\tau_p$ satisfying the conditions $\Omega\tau_p \ll 1$ and $\varkappa \tau_p \ll 1$. Since usually $\gamma \lesssim \varkappa$ the condition $\gamma \tau_p \ll 1$ is fulfilled as well. As for the product  $g\tau_p$, its value is left arbitrary. The process of spin initialization or control of a resident electron by short circularly-polarized pulses can conveniently be divided into three stages corresponding, in accordance with the inequalities~\eqref{b:field}, to three different time scales: (i) on the time scale of $t \sim \tau_p$ where both the magnetic field and relaxation processes are unimportant and the action of a pump/control pulse can be described in the framework of the Hamiltonian~\eqref{ham} with $\Omega=0$, (ii) on the time scale $\tau_p \ll t \sim 1/\Omega \ll 1/\varkappa$ one should take into account the magnetic field in Eq.~\eqref{ham}, and (iii) for $1/\Omega \ll t \sim 1/\varkappa$ the photon decay becomes important. Such separation of time scales strongly simplifies the calculations. To illustrate the approach, we consider the spin dynamics under the excitation by $\sigma^+$ polarized pulses. Similarly to the pump-probe experiments on single QDs and QD arrays without a cavity \cite{yugova09}, the repetition period of the pump pulses is assumed to be longer than the damping times $\gamma^{-1}$ and $\varkappa^{-1}$. Hence the accumulated photons have the time to escape the cavity and the system returns to the ground state before the arrival of the next optical pulse. Therefore, it is sufficient to trace the behavior of the system following a single optical pulse.

In addition, we point out that for a circularly-polarized excitation the system evolution is simplified. If only $\sigma^+$ photons are injected into the cavity, i.e., $\mathcal{E}_-(t)=0$, the system evolution is limited to the subset of states corresponding to $m_-=0$, and the states with spin $\Downarrow$ remain unexcited. With explicit notations, we thus use a basis formed by the states $\left|{\downarrow},m\times{\sigma^+}\right\rangle$, $\left|{\uparrow},m\times{\sigma^+}\right\rangle$ and $\left|{\Uparrow},m\times{\sigma^+}\right\rangle$, corresponding to $m=0, 1,\ldots$ $\sigma^+$ photons and no $\sigma^-$ photons.

Let the resident electron, before the pulse arrival ($t \to - \infty$) be in the state given by the spinor components $\psi_{1/2}, \psi_{-1/2}$ normalized to the unity. In the above-mentioned notations, the total initial system state is thus given by $\psi_{1/2}\left|{\uparrow},0\times{\sigma^+}\right\rangle + \psi_{-1/2}\left|\downarrow,0\times\sigma^+\right\rangle$. In the first stage, i.e. at the direct action of the pump pulse ($\mathcal E_+\ne 0$), the system may be described by a time-dependent wavefunction given by
\begin{eqnarray}
\label{wavefunction}
 \Psi  &=&\sum_{m=0}^{\infty} \left\{ C^-_m(t)\left|\uparrow,m\times{\sigma^+}\right\rangle e^{-{\rm i}m\omega_c t} \right. \nonumber
 \\ && +\
 C^+_m{(t)} \left|\Uparrow,m\times{\sigma^+}\right\rangle e^{-{\rm i}(m+1)\omega_c t} \hspace{0.8 cm} \nonumber \\
 && \left. +~ D_m(t) \left|\downarrow,m\times\sigma^+\right> e^{-{\rm i}m\omega_c t} \right\}\:, \hspace{1.6 cm}
\end{eqnarray}
In this equation the coefficients $C^-_m(t)$ and $C^+_m(t)$ correspond to states $\left|\uparrow,m\times{\sigma^+}\right\rangle$ and $\left|\Uparrow,m\times{\sigma^+}\right\rangle$, respectively. Note that the states  $\left|\uparrow,m\times{\sigma^+}\right\rangle$ and $\left|\Uparrow,m-1\times{\sigma^+}\right\rangle$ are coupled through optical electron-trion transition, i.e., via the light-matter interaction. The coefficients $D_m(t)$ correspond to the uncoupled states $\left|\downarrow,m\times{\sigma^+}\right\rangle$, for which no interaction occurs between the electron spin $\downarrow$ and the $\sigma^+$ photons. These coefficients satisfy the initial conditions: $C^-_0(-\infty) = \psi_{1/2}, \quad D_0(-\infty) = \psi_{-1/2}$ and all other coefficients vanishing. The coefficients $C^\pm_m(t)$ obey the following infinite set of coupled differential equations
\begin{widetext}
\begin{subequations}
\label{C}
\begin{align}   \label{Cm}
  {\rm i}\dot{C}^-_m&=\sqrt{m}gC^+_{m-1}+\mathcal E_+(t)\left(\sqrt{m}e^{{\rm i}\delta t}C^-_{m-1}+\sqrt{m+1}e^{-{\rm i}\delta t}C^-_{m+1}\right)\:,\\
  {\rm i}\dot{C}^+_m&=\Delta C^+_m + \sqrt{m+1}gC^-_{m+1}+\mathcal E_+(t)\left(\sqrt{m}e^{{\rm i}\delta t}C^+_{m-1}+\sqrt{m+1}e^{-{\rm i}\delta t}C^+_{m+1}\right)\:.
\end{align}
\end{subequations}
\end{widetext}
Here $\delta=\omega_c-\omega$ and $\Delta=\omega_0-\omega_c$ are the detunings, respectively, between the cavity-mode frequency and the pulse carrier frequency, and between the quantum-dot and cavity-mode resonant frequencies, and the time variable in $C_m^\pm(t)$ and $D_m(t)$ is  omitted for brevity. Equations of motion for $D_m$ can be obtained from \eqref{Cm} by setting $g=0$:
\begin{eqnarray} \label{D}
{\rm i} \dot{D}_m &=& \mathcal E_+(t) \left( \sqrt{m}e^{{\rm i}\delta t} D_{m-1} \right. \nonumber \\
&& \left.  +\  \sqrt{m+1} e^{-{\rm i}\delta t}D_{m+1} \right) \:.
\end{eqnarray} 
The latter permits an analytical solution for the real $\mathcal E_+(t)$ as follows~\cite{glauber63}
\begin{equation}
\label{D:sol}
D_m = \psi_{-1/2}e^{{\rm i}\varphi} \frac{(-{\rm i}{\alpha})^m}{\sqrt{m!}}e^{-|{\alpha}|^2/2}\:,
\end{equation}
where 
\begin{equation}
\alpha(t)=\int\limits_{-\infty}^t\mathcal E_+(t')e^{{\rm i}\delta t'}dt',
\label{alpha}
\end{equation} 
and 
\[
\varphi(t)=\int\limits_{-\infty}^t\d t'\mathcal E_+(t')\int\limits_0^\infty\d\tau\mathcal E_+(t'-\tau)\sin(\delta\tau).
\] 
According to Eq.~\eqref{wavefunction}, after the pulse the system is described by a coherent superposition of the electron, trion and multiphoton states. The dynamics under a $\sigma^-$ polarized pulse can be described similarly. It is worth stressing that the statistics of uncoupled ladder of states, related with $|D_m|^2$, is described by the coherent Glauber distribution inherent for the classical light~\cite{glauber63}. By contrast, due to the presence of two-level system non-linearity the coefficients $|C_m^\pm|^2$ in Eqs.~(\ref{C}) do not obey coherent distribution~\cite{poshakinskii}. Therefore, to describe spin initialization, control and detection the field quantization in the cavity should be explicitly taken into account.

At the second stage, started after the short pulse has passed and lasting during the time $\sim\Omega^{-1}$, the magnetic field comes into play and
the Shr\"odinger equation reduces to the set
\begin{subequations}
\label{B-eq}
\begin{align}
 &{\rm i}\dot{C}^-_{m}=\sqrt{m}gC^+_{m-1}+\frac{\Omega}{2}D_m \:,\\
  &{\rm i}\dot{C}^+_m=\Delta C^+_{m}+\sqrt{m+1}gC^-_{m+1},\\
  &{\rm i}\dot{D}_m=\frac{\Omega}{2}C^-_{m} \:.
\end{align}
\end{subequations}
A direct consequence of the above equations is that under the assumption $\Omega \ll g$, the transverse magnetic field mixes, in the first order, only the ground (no-photon/trion) states $\left|\uparrow,0\times\sigma^+\right\rangle$ and $\left|\downarrow,0\times\sigma^+\right\rangle$. The set~(\ref{B-eq}) is decomposed into a pair of closed equations 
\begin{equation} \label{C0D0}
{\rm i}\dot{C}^-_{0}=\frac{\Omega}{2}D_0\:,\:{\rm i}\dot{D}_0=\frac{\Omega}{2}C^-_{0} 
\end{equation}
describing the electron spin in the no-photon ground state with $m=0$, and other equations
\begin{eqnarray}
&& {\rm i}\dot{C}^-_m=\sqrt{m}gC^+_{m-1} \hspace{3 mm} (m > 0)\:, \nonumber \\ &&{\rm i}\dot{C}^+_m = \Delta C^+_m+\sqrt{m+1}g C^-_{m+1} \hspace{2 mm} (m \geq 0)\:, \nonumber\\
 &&{\rm i}\dot{D}_m= 0 \hspace{3 mm} (m > 0)\:.
\end{eqnarray}
The initial conditions for them are given by the limiting values $C^{\pm}_m(+\infty), D_m(+\infty)$ found in the first stage. 
The components of the electron spin ${\bm S=(S_x,S_y,S_z)}$ in the ground bare states $\left|\uparrow,0\times\sigma^+\right\rangle$ and $\left|\downarrow,0\times\sigma^+\right\rangle$ are determined by
\begin{subequations}
\label{Sxyz}
\begin{align}
S_x &= \frac12 \left[C^-_0D^*_0+{C^-_0}^*D_0 \right]\:,\\
S_y &= \frac{\rm i}{2} \left[ C^-_0D^*_0-{C^-_0}^*D_0\right]\:,\\
S_z &= \frac12 \left[|C^-_0|^2-|D_0|^2\right]\:. \label{S0z}
\end{align}
\end{subequations}
The magnetic field ${\bm B} \parallel x$ induces the Larmor spin precession in the $(yz)$ plane
\begin{subequations}
\label{precession0}
\begin{align}
S_z(t) &= S_z^{(0)} \cos{\Omega t} + S_y^{(0)} \sin{\Omega t} \:, \label{pr10}\\
S_y(t) &= S_y^{(0)} \cos{\Omega t} - S_z^{(0)} \sin{\Omega t} \:, \label{pr20}\\
S_x(t) &= S_x^{(0)}\:. \label{pr30}
\end{align}
\end{subequations}
where the initial components $S_i^{(0)}$ ($i=x,y,z$) are related by Eqs.~(\ref{Sxyz}) with the first-stage coefficients $C^-_0(+\infty)$ and $D_0(+\infty)$. Here we stress that the amplitude of the Larmor spin precession is governed only by the populations and coherences for the two ground states $\left|\uparrow,0\times\sigma^+\right\rangle$ and $\left|\downarrow,0\times\sigma^+\right\rangle$.

On the third stage, $t\sim \varkappa^{-1}$, irreversible relaxation processes should be included into the analysis. The trion decay and the photon escape return the system to the ground state. However, for high enough magnetic field, $\Omega \gg \varkappa$, the electron returning from the states $\left|{\uparrow},m \times{\sigma^+}\right\rangle$, $\left|{\downarrow},m \times{\sigma^+}\right\rangle$ with $m > 0$ and the trion states into the ground states makes no contribution to the rotating spin ${\bm S}(t)$. Indeed, if $\Omega \gg \varkappa$ the states $\left|\uparrow,0\times\sigma^+\right\rangle$ and $\left|\downarrow,0\times\sigma^+\right\rangle$ experience a Larmor rotation which is much faster than the typical time intervals between the successive photon or trion decay events. These events incoherently contribute to increase the population of either $\left|\uparrow,0\times\sigma^+\right\rangle$ or $\left|\downarrow,0\times\sigma^+\right\rangle$ state. Thus, they leave both the direction and magnitude of the spin vector ${\bm S}(t)$ unchanged, as confirmed by a numerical calculation carried out within the density matrix formalism for the condition $\Omega \gg \varkappa$. Qualitatively, this can be understood by transforming to the coordinate system rotating with the Larmor frequency around the field ${\bm B}$: In this system the ground-state electron spin is fixed whereas the direction of the electron spin incoming from the decaying excited states is changing with the frequency $\Omega$. As a result, the spin dynamics is not modified during the third stage and closely follows Eqs~\eqref{precession0} if $\Omega \gg \varkappa$ (the situation where $\kappa$ is comparable to $\Omega$ is described below and in Appendix~\ref{sec:reson}).

Finally, when considering timescales $t \gg \varkappa^{-1}, \gamma^{-1}$, one also has to take into account the spin relaxation for the resident electron. For these timescales
the spin dynamics is modified, as compared to Eqs.~(\ref{pr10}) and (\ref{pr20}), by 
\begin{subequations}
\label{precession}
\begin{align}
S_z(t) &= \left( S_z^{(0)} \cos{\Omega t} + S_y^{(0)} \sin{\Omega t} \right){\rm e}^{-t/\tau_s}\:, \label{pr1}\\
S_y(t) &= \left(S_y^{(0)} \cos{\Omega t} - S_z^{(0)} \sin{\Omega t} \right) {\rm e}^{-t/\tau_s}\:, \label{pr2}\\
S_x(t) &= S_x^{(0)} {\rm e}^{-t/\tau_s} \:,
\end{align}
\end{subequations}
where $\tau_s$ is the phenomenological spin relaxation time of resident electron assumed to exceed by far $\Omega^{-1}$, $\varkappa^{-1}$, and $\gamma^{-1}$. Equations \eqref{precession} together with (\ref{Sxyz}) fully describe the resident-electron spin dynamics induced by the short circularly-polarized pulse. We also point out that, in typical experiments, periodic trains of pulses are used, with a repetition rate of the order or even much higher than the spin relaxation rate. The effect of periodic pump pulses is discussed in Appendix~\ref{sec:train}, showing that very efficient spin initialization can be obtained even when considering the incomplete spin initialization induced by a single pump pulse and a relaxation of the spin coherence between two consecutive pulses. Using this formalism we address below the spin coherence initialization and spin control effects.

\subsection{Spin coherence initialization}

\begin{figure}[t]
 \includegraphics[width=0.99\linewidth]{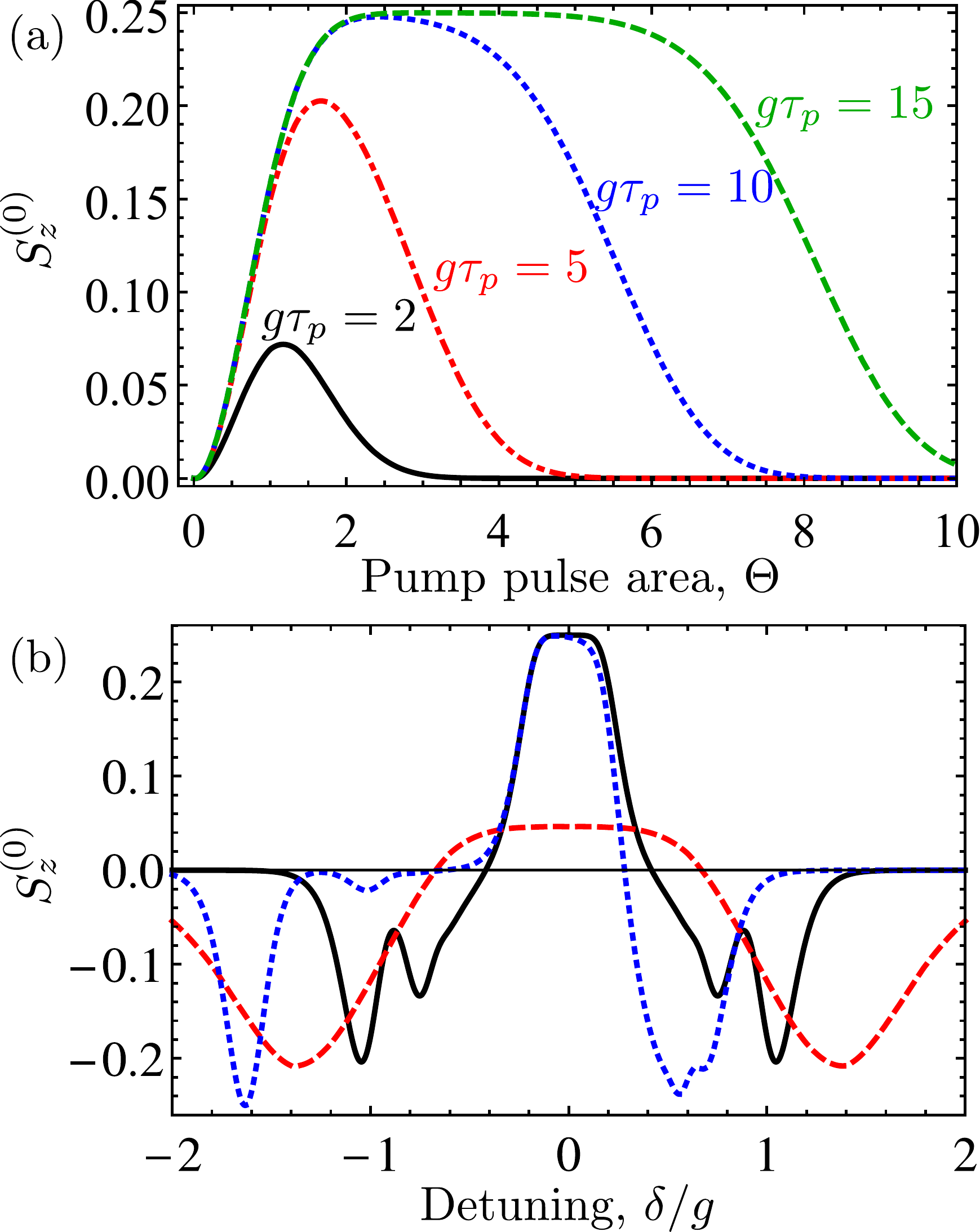}
\caption{(a) Spin $S_z^{(0)}$ of a resident electron after a single $\sigma^+$ short pump pulse vs. the pulse area under the resonant excitation
of the microcavity tuned to resonance with the trion transition, i.e. for $\omega=\omega_0=\omega_c$. Different curves correspond to different pump pulse durations indicated at each curve. (b) The electron spin $S_z^{(0)}$ as a function of the pump-pulse detuning $\delta$ at the fixed pulse area $\Theta = 3$ calculated after Eqs.~\eqref{C}, \eqref{D} and \eqref{Sxyz} for the following parameters: $\Delta=0$ and $g\tau_p=15$ (solid curve), $\Delta=0$ and $g\tau_p=4$ (dashed curve), $\Delta=g$ and $g\tau_p=15$ (dotted curve).}
\label{fig:pump}
\end{figure}

Figure~\ref{fig:pump} demonstrates the electron spin polarization $S_z^{(0)}$ induced by a short $\sigma^+$-polarized pump pulse as (a) a function of pump pulse area introduced by
\begin{equation}
\label{area}
\Theta = \int_{-\infty}^\infty\mathcal E_+(t')dt'\:,
\end{equation}
at the double resonant condition $\omega=\omega_0 = \omega_c$, and (b) a function of the detuning between the pump pulse carrier frequency $\omega$ and the cavity resonant frequency $\omega_c$. To model the initially unpolarized electron we used the spinor components $\psi_{1/2}, \psi_{-1/2}$ of equal absolute values, $|\psi_{1/2}| = |\psi_{-1/2}|$, and averaged $S_z^{(0)}$ over their relative phase. 
The envelope function of the pump pulse is taken in the form of Rosen \& Zener pulse{~\cite{PhysRev.40.502}}
\[
\mathcal E_+(t)=\frac{\Theta}{\tau_p\cosh\left(\pi t/\tau_p\right)}\:.
\]

As follows from Fig.~\ref{fig:pump}(a), at small values of $\Theta$ (weak pulses) the electron spin polarization increases proportionally to $\Theta^2$ or to the first power of the light intensity, as expected from general considerations. With the further increase in $\Theta$ it reaches a maximum and then slopes down. The decrease in polarization is related to depopulation of the ground states 
$\left|{\uparrow},0\times{\sigma^+}\right\rangle$ and $\left|{\downarrow},0\times{\sigma^+}\right\rangle$: for optical pulses of high intensity the coefficients $C^-_0(+\infty), D_0(+\infty)\to 0$ and, according to Eqs.~\eqref{Sxyz}, ${\bm S}(0) \to 0$. This is because spectral wings of high intensity pulse induce non-resonant transitions, see below. It is worth to emphasize that the $\Theta$-dependence does not show any oscillations, in contrast to the Rabi cycle inherent to two-level systems~\cite{ll3_eng,yugova09}. This can be understood taking into account an infinite number of strongly-coupled excited states in the system under consideration, see detailed discussion on Rabi effect in Appendix~\ref{sec:Rabi}. Furthermore, a single pump pulse cannot polarize electron spin by more than $S_z=0.25$ if initially the electron is unpolarized. In the optimal case, indeed, the pulse will fully transfer one of the spin components to the excited states, and leave unaffected the other spin component, i.e. $|D_0(+\infty)|^2=0$ and $|C^-_0(+\infty)|^2={1}/{2}$ right after the pump pulse. In such a case, the trion and photon decay processes will equally transfer the populations of the excited states towards the two ground states.
This corresponds to a maximal polarization degree $\rho_z = 2 S_z \leq 50\%$ if the spin is initially unpolarized before the pulse; however, as shown in Appendix~\ref{sec:train}, a train of synchronized pump pulses can result in the complete spin polarization of the resident carrier~\cite{yugova09}.

\begin{figure}[t]
\includegraphics[width=\linewidth]{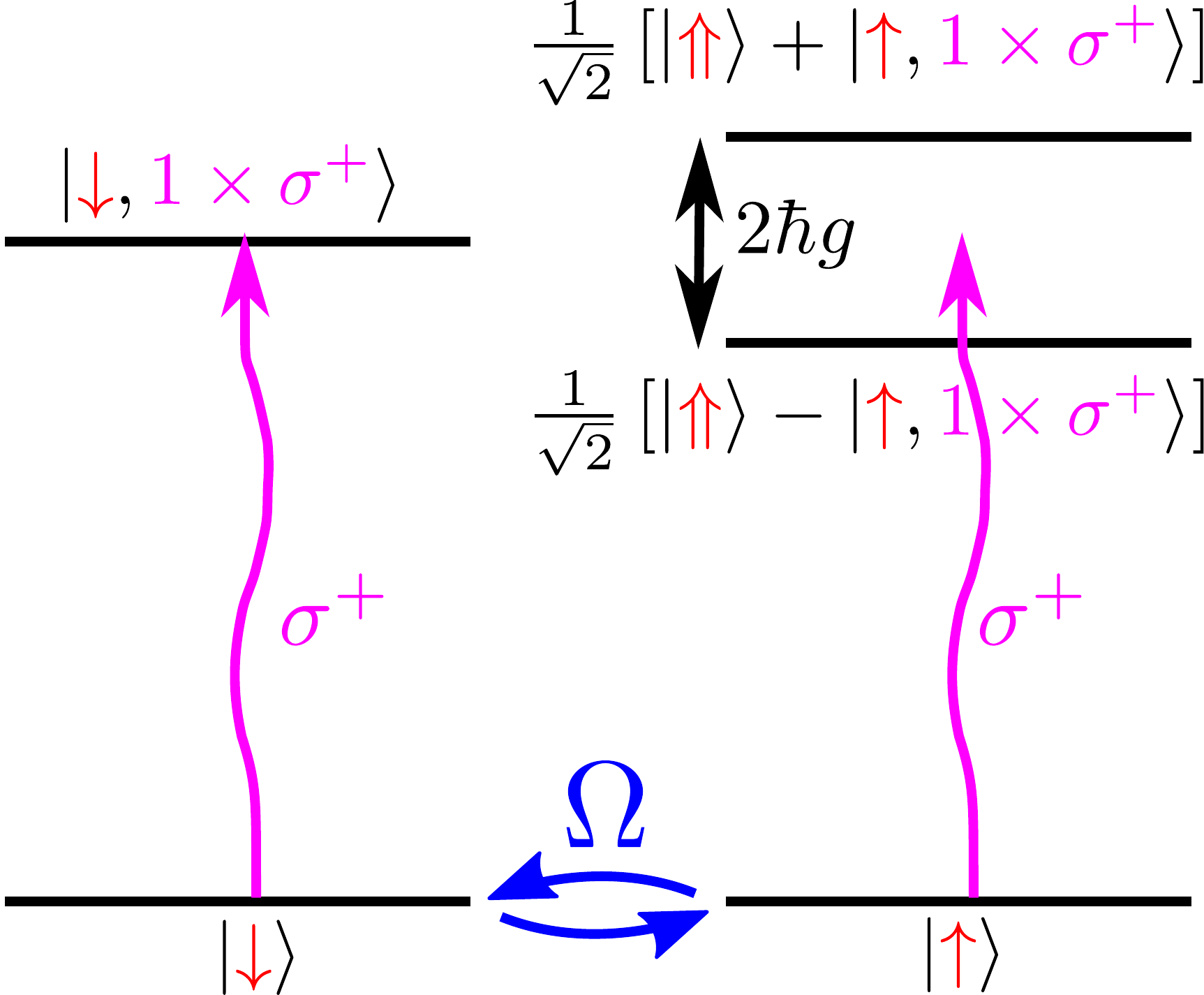}
\caption{ Scheme of the energy levels and selection rules for transitions involving a single photon. Double arrows $\Uparrow$, $\Downarrow$ denote heavy-hole-in-trion spin, single arrows $\uparrow$, $\downarrow$ denote resident electron spin. The transitions between the states after injection of single $\sigma^+$ photon are shown by wavy arrow. The polariton modes are split by $2\hbar g$ and a weak external magnetic field $B$ acts only on the ground electron states, see main text for details.}
\label{fig:scheme:tr}
\end{figure}

We now turn to the spectral features of this spin initialization procedure, which can be understood by considering the eigenstates of the system in the absence of any external field (i.e. when $\mathcal E_\pm(t)$ and $\Omega=0$). Indeed, the condition $g \gg \varkappa,\gamma$ corresponds to a strong coupling regime with a clear spectral separation between the polaritonic states induced by the light-matter coupling. In the absence of $\sigma^-$ photons, the eigenstates we have to consider in our system are the ground states  $\left|\downarrow,0\times{\sigma^+}\right\rangle$ and  $\left|\uparrow,0\times{\sigma^+}\right\rangle$, the uncoupled states $\left|\downarrow,m\times{\sigma^+}\right\rangle$ with $m>0$, and the polaritonic eigenstates of the form 
\begin{equation}
\label{polaritons}
\frac{1}{\sqrt{2}}\left[ \left|\uparrow,m\times{\sigma^+}\right\rangle \pm \left|\Uparrow,(m-1)\times{\sigma^+}\right\rangle\right],
\end{equation}
also with $m>0$. While the energies $E^{0}_{m}$ of the uncoupled eigenstates $\left|\downarrow,m\times{\sigma^+}\right\rangle$ are simply given by
\begin{equation}
\label{spectrum_uncoupled}
E^{0}_{m} = m \hbar  \omega_0, \quad (m=1, 2, \ldots)\:,
\end{equation}
the energies $E^{\pm}_{m}$ of the polaritonic eigenstates, Eq.~\eqref{polaritons}, are given by the Jaynes-Cummings formulae~\cite{JC}:
\begin{equation}
\label{spectrum}
E^{\pm}_{m} = E^{0}_{m} \pm \sqrt{m} \hbar g, \quad (m=1, 2, \ldots)\:.
\end{equation}
Figure~\ref{fig:scheme:tr} presents the energy spectrum of this system for the two ground states, and for the first three excited states, illustrating that the energy degeneracy is lifted only for the excited states. This is the reason why, when a magnetic field is turned on in the limit $\Omega\ll g$, this leads to an efficient mixing between the ground states but not between the excited states. Figure~\ref{fig:scheme:tr} also demonstrates, why, in the results of Fig.~\ref{fig:pump}(a), efficient spin initialization is obtained when the pump pulse is longer than the inverse coupling constant $g^{-1}$, i.e. $g \tau_p \gg 1$. Indeed, the wider the pulse spectrum, the weaker the rate of resonant transition and, in addition, the spectrally wide pulses give rise to transitions involving both spin-up and spin-down electron states. For $\hbar g=500~\mu$eV, the condition $g\tau_p > 1$ corresponds to $\tau_p > 1$ ps.

Under the double resonance condition, $\omega = \omega_c = \omega_0$, the right-handed circularly-polarized pump pulse orients the electron spin in the positive $z$ direction, as shown in Fig.~\ref{fig:pump}(a) and in Fig. \ref{fig:pump}(b) for zero detuning. To explain this sign of $S_z^{(0)}$ we refer to the wavy arrows in Fig.~\ref{fig:scheme:tr}: for a large enough coupling constant $g$, the carrier frequency $\omega$ is out of resonance with the split excited states (\ref{polaritons}) and in resonance with the transitions 
$$
\left| \downarrow, m \times \sigma^+ \right> \to \left| \downarrow, (m + 1) \times \sigma^+ \right>\:.
$$ 
As a result the $\sigma^+$ pulse causes a reduction of the coefficient $D_0(+\infty)$, exciting and therefore disordering the electron spin-down state. It is clear that one can invert the spin polarization by detuning the pulse frequency from $\omega_c = \omega_0$ to one of the polariton transitions, thus causing $C^-_0(+\infty)\to 0$ and $S_z^{(0)}<0$, in agreement with the behavior of the solid curve in Fig.~\ref{fig:pump}(b). As mentioned above, the high-intensity pulse results in the reduction of both $C_0^-$ and $D_0$ since the power of spectral wings increases with the pulse area. An extra feature in the curve is related to the resonant two-photon excitation of the $m=2$ polaritonic states, at energies given by $E^{\pm}_{2}$ in Eq. \eqref{spectrum}: For this process the resonance condition is $2 \omega = 2 \omega_c \pm \sqrt{2} g$ or $|\delta|\approx g/\sqrt{2}$ being a signature of the second rung in the Jaynes-Cummings ladder. For the larger detunings, $|\delta| > g$, the optical transition efficiency decreases and the electron spin orientation is suppressed, similarly to the above-mentioned effect of pulse duration shortening, compare also the solid and dashed curves in Fig.~\ref{fig:pump}(b). The detuning $\Delta = g$ between the trion and cavity-mode resonances shifts leftwards the dips related to the spin-up polaritons, as seen from the dotted curve in Fig.~\ref{fig:pump}(b).

\begin{figure}[t]
\includegraphics[width=0.99\linewidth]{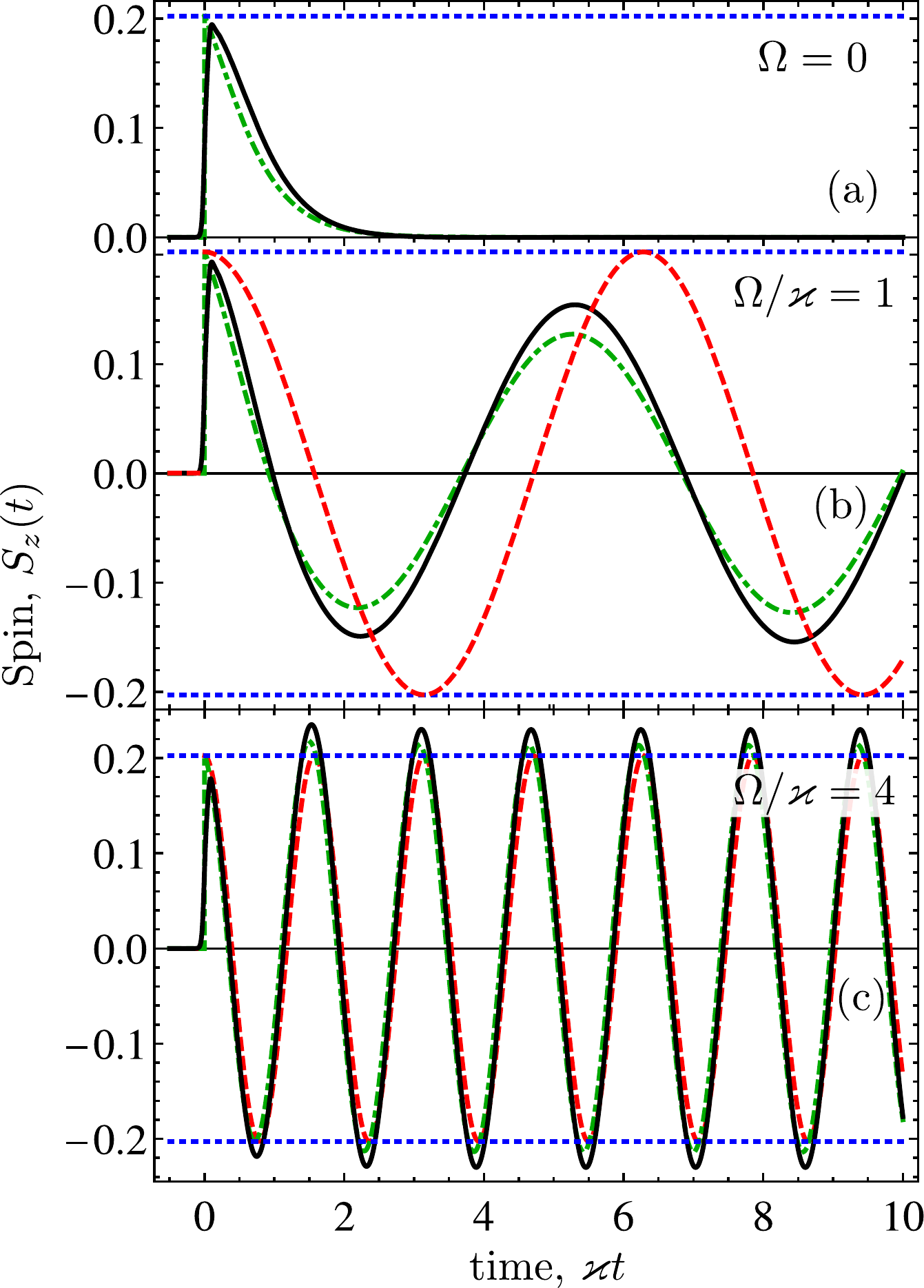}
\caption{The spin beats calculated for different Larmor frequencies $\Omega=0,\varkappa,4\varkappa$ as indicated in the panels. The black solid curves are calculated in the density matrix formalism using Eqs.~\eqref{density:m} and \eqref{Lindblad}. The red dashed curves are calculated for $t>0$ after Eqs.~\eqref{precession0} with the amplitude determined by Eqs.~\eqref{C}, \eqref{D} and \eqref{Sxyz} (indicated by blue dotted lines). The green dash-dotted line is calculated using the resonant approximation after Eqs.~\eqref{reson}. The parameters of the calculation are $\Delta=\delta=0$, $\Theta=1.7$, $g \tau_p=5$, $\varkappa=g/50$ and $\gamma=g/250$.}\label{fig:beats}
\end{figure}

The results obtained within the simplified three-stage model presented above are corroborated by the numerical calculations performed within the full density matrix approach. This approach was also applied to calculate explicitly the spin beats after Eqs.~\eqref{density:m} and \eqref{Lindblad}. 
The results are presented in Fig.~\ref{fig:beats} by black solid lines together with the predictions of the simplified three-stage model (red/dashed lines) for three different magnitudes of magnetic field. The parameters of calculations are presented in the figure caption and, as above, the initial condition corresponds to the unpolarized electron. In the case of $\Omega=0$ the long-living spin beats (at $t\gtrsim 1/\varkappa$) are not excited, because in the absence of both magnetic field and spin flip processes, the electron spin returning from the excited states exactly compensates the spin generated by the pump pulse.

Accordingly, an increase of $\Omega$ leads to a non-zero amplitude of the long-living spin beats. For large enough magnetic field, where $\Omega \gg \varkappa$ and the inequalities \eqref{b:field} are fulfilled, the spin beats calculated in the density matrix formalism and in the three-stage model expectedly coincide. A more elaborate model of spin coherence initialization outlined in Appendix~\ref{sec:reson} allows us to describe spin coherence excitation even for $\Omega \lesssim \varkappa$, corresponding results of calculations are presented by green dash-dotted lines in the Fig.~\ref{fig:beats}.

\begin{figure}[t]
\includegraphics[width=0.99\linewidth]{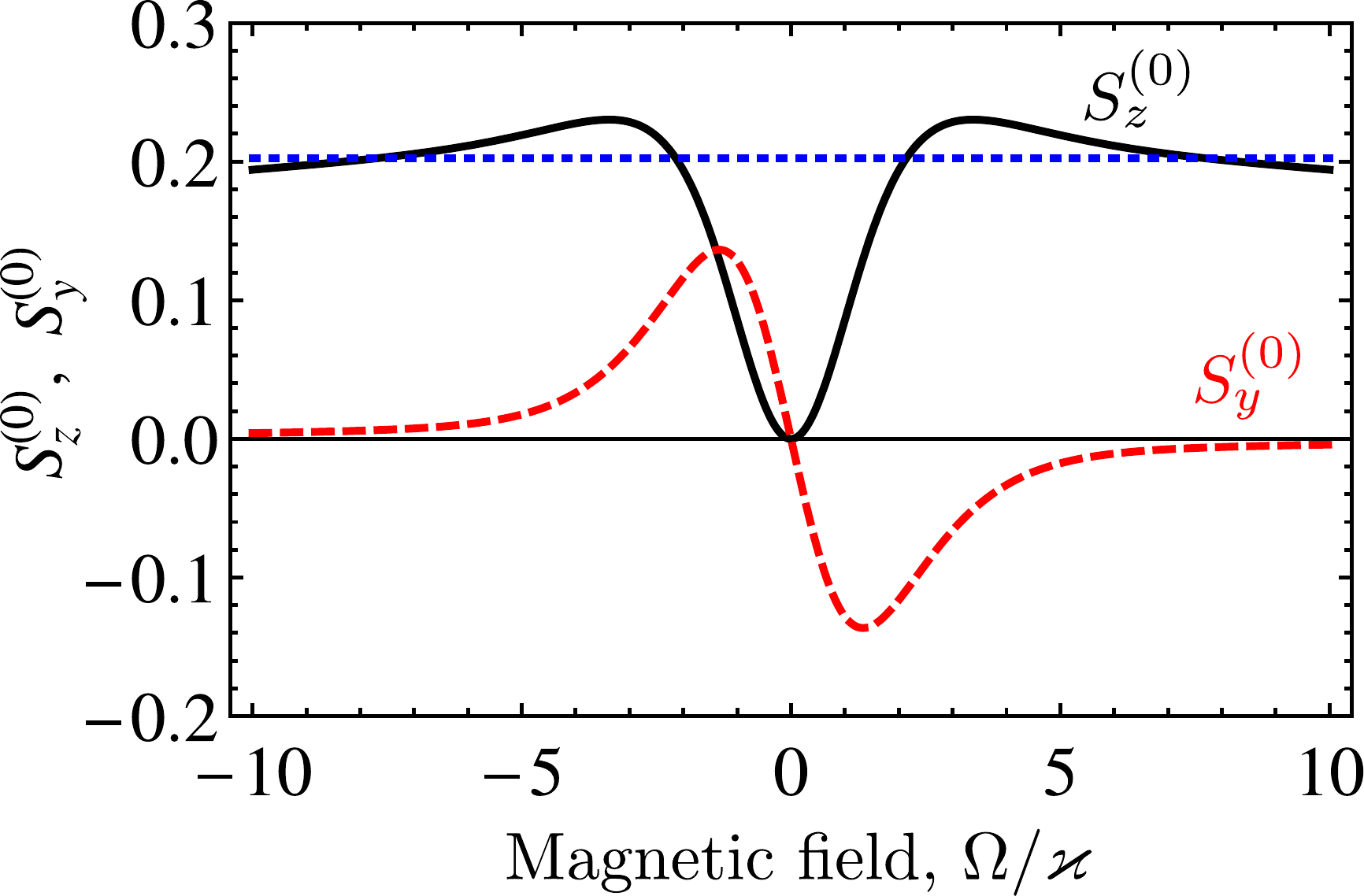}
\caption{Magnetic-field-induced variation of the spin polarizations $S_z^{(0)}$ and $S_y^{(0)}$ calculated in the density matrix formalism using Eqs.~\eqref{density:m} and \eqref{Lindblad}, black solid and red dashed curves respectively, with the same parameters as in Fig.~\ref{fig:beats} and $T_{max}=10/\varkappa$.}\label{fig:magnetic}
\end{figure}

The dependence of the spin beats on the ratio $\Omega/\varkappa$ is also illustrated in Fig.~\ref{fig:magnetic}. Fitting the numerically-calculated spin beats with Eqs.~\eqref{precession}, at the time scale ranging from $\tau_p$ to some value $T_{max}$ chosen in a way that $1/\varkappa \ll T_{max} \ll \tau_s$, allows us to determine the initial spin components $S_z^{(0)}$ and $S_y^{(0)}$ in Eqs.~\eqref{precession}, and compare them to the prediction of Eq.~\eqref{S0z}. The deduced values of $S_z^{(0)}$ and $S_y^{(0)}$ are shown in Fig.~\ref{fig:magnetic} as functions of the magnetic field. At small magnetic fields, $\Omega \lesssim \varkappa$, where the three-stage model is inapplicable the density matrix calculation predicts the quadratic and linear dependencies of $S_z^{(0)}$ and $S_y^{(0)}$ on the magnetic field. At the higher fields, $\Omega \gg \varkappa$, the electron spin $S_y^{(0)}$ vanishes and $S_{z}^{(0)}$ saturates at a value given by Eq.~\eqref{S0z}. A slow decrease of $S_{z}^{(0)}$ in this region is related with the increasing ratio $\Omega/g$ assumed to be small in the Schr\"{o}dinger equation approach.  These properties are similar to those in the previously-studied spin initialization in quantum wells and QDs without a cavity~\cite{glazov:review,yugova12,bayer_long}. Hence, the comparison of the developed 
three-stage model and full density matrix approach demonstrates that under the condition $\Omega \gg \varkappa$ one can indeed neglect the contribution of excited states to the spin beats.

\subsection{Spin coherence control}

\begin{figure}[t]
\includegraphics[width=0.99\linewidth]{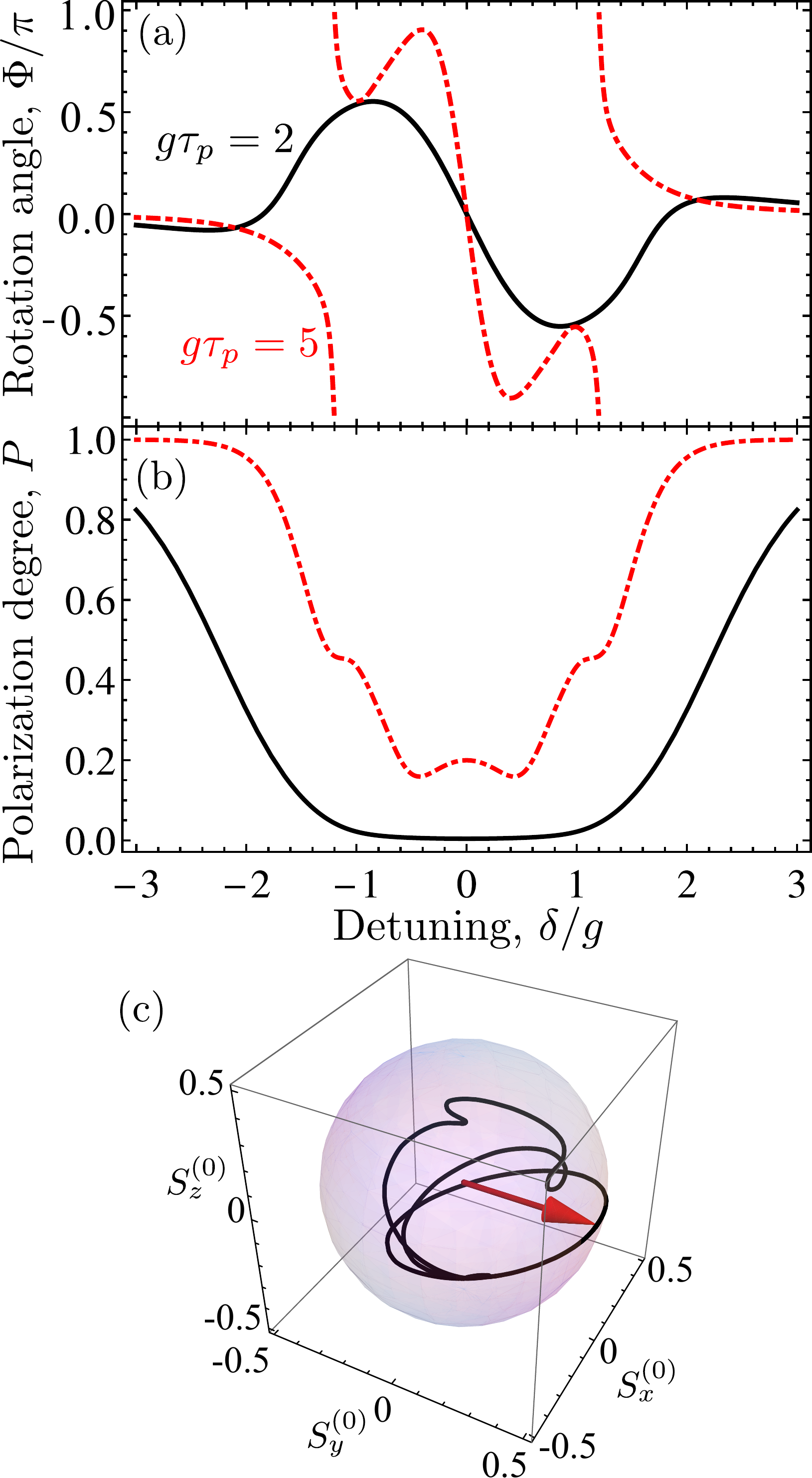}
\caption{(a) Spin rotation angle $\Phi$ as a function of the detuning between the cavity resonance and carrier frequency of control pulse for $\Theta=3$ and two control pulse durations {indicated at the curves}. (b) Electron spin polarization degree after the control pulse arrival calculated for the same parameters as in (a). (c) Trajectory of spin vector $\bm S$ under variation of detuning $\omega - \omega_0$ from $-5g$ to $+5g$ for $\Theta=3$ and $g\tau_p=15$.}\label{fig:rot}
\end{figure}

Equations~\eqref{C}, \eqref{D:sol}, and {\eqref{precession0}} describe also the coherent spin manipulation by short circularly polarized pulses. To address such a situation we assume that shortly before the control pulse arrival the resident electron spin is aligned along $y$-axis, yielding
\[
\psi_{1/2} = {1/\sqrt{2},\quad} \psi_{-1/2} ={\mathrm i}/\sqrt{2}.
\]
In the absence of the control pulse, this situation would simply correspond to the spin precession described by Eqs. \eqref{precession0} with initial conditions $S_x^{(0)}=0$, $S_y^{(0)}=1/2$, and $S_z^{(0)}=0$. After the control pulse, the spin orientation will still be described by Eqs \eqref{precession0}, but with modified values of the coefficients $S_x^{(0)}$, $S_y^{(0)}$, and $S_z^{(0)}$. Indeed, similarly to the previously studied situation of the QD without the cavity~\cite{carter:167403,yugova09,spherical-dots,glazov:review} the circularly polarized pulse induces spin rotation in the $(xy)$ plane. The spin rotation angle induced by the control pulse is defined from the phase of the spin beats in Eqs.~\eqref{precession0}: $\Phi=\arg\left(S_y^{(0)}-i S_x^{(0)}\right)$. This rotation angle $\Phi$, calculated in the three-stage model, is presented in Fig.~\ref{fig:rot}(a) as a function of the detuning $\delta$ between the carrier frequency and QD/cavity resonances (here and below $\omega_0=\omega_c$). Here, as before, the conditions $g\gg \Omega \gg \varkappa,\gamma$ were assumed to be satisfied.

Figure~\ref{fig:rot}(a) shows that spectrally narrow control pulses can rotate the resident electron spin in the $(xy)$ plane by an arbitrary angle. Particularly, as demonstrated with the dashed curve, calculated for $g\tau_p=5$, the spin rotation angle in the plane can reach $2\pi$. For somewhat shorter pulses $g\tau_p=2$ the spin rotation angle is still significant.

We now show that besides spin rotation, the control pulse also generally induces a spin depolarization, which, however, can remain limited under proper excitation conditions. In the optical control situation, indeed, one specifically starts with a situation where $S_z=0$. Because the two ground states are populated, the control pulse will populate at least one of the excited states, thus leading to a partially polarized spin right after the control pulse. The spin will then remain partially polarized after the depopulation of the excited states, on the time scale of $t\sim 1/\varkappa^{-1}$. This depolarizing effect is similar to the one which is obtained with a QD without a microcavity, whenever a trion population is created by the control pulse: the trion decay at random times then leads to an unavoidable spin depolarization. In order to limit this depolarization, a well known technique in QDs without microcavities consists in using specific values of the pulse area $\Theta=2\pi n$, where $n$ is an integer~\cite{A.Greilich07212006,glazov:review}, so that the trion state populations remain negligible after the control pulse. A similar technique can be applied in the present case, by applying a control pulse which reduces the population of the excited states right after the pulse, see Appendix~\ref{sec:Rabi} where the conditions for the Rabi cycle are analyzed.

In order to illustrate the effect of the control pulse, the degree of spin polarization in the strong coupling regime is plotted in Fig.~\ref{fig:rot}(b) as a function of the detuning. The  depolarization effect is present for any values of $\Theta$ and $\delta$, but can remain moderate for given values of the pulse area, as for example with $g\tau_p=5$, $\Theta=3$, and $\delta/g = \pm 1.4$ in Fig.~\ref{fig:rot}(b).

The panel (c) of Fig.~\ref{fig:rot} visualizes the spin control and presents the possible spin polarizations of the QD after the control pulse arrival for different detunings $\delta$ provided the spin is initially oriented along $y$ axis. Note that this trajectory can be simpler or even more complex depending on the pulse power and duration. The longer is the pulse the more polarization can be kept for the given rotation angle. For example, calculation shows that taking $g\tau_p=15$ and $\Theta=3$ one can rotate spin by $\Phi=\pi/2$ with small depolarization of about 15\%.

We also point out that this possible depolarization does not contradict the fact that, in the formerly described spin initialization process, perfect spin polarization can be achieved using a train of circularly polarized optical pulses. Indeed, as shown in Appendix~\ref{sec:train}, under the double resonance condition $\omega=\omega_c=\omega_0$, and with a proper synchronisation of the pulses, one can reach a situation where $S_z=1/2$ before and after a pulse arrival: in such a case the only state which is populated at the pulse arrival is the ground state $|\uparrow,0,0\rangle$. This states remains unaffected by the circularly polarized incoming pulse, which therefore does not induce neither any spin depolarization nor spin rotation.

\section{Spin coherence detection}\label{sec:det}

The spin polarization is detected by a linearly polarized probe pulse. In the strong coupling regime the Faraday and Kerr effects induced by a single electron spin become giant, since one circular component of the probe pulse can be fully transmitted through the cavity, while the other one can be fully reflected~\cite{PhysRevB.78.085307,Arnold2015}. It follows from Refs.~\cite{PhysRevB.78.085307,milburn} that in the limit of small amplitude of the probe beam, $|\mathcal E_i|/\varkappa \ll 1$, the reflection coefficients, $r_\pm$, of the monochromatic $\sigma^{\pm}$ polarized light from the microcavity can be presented as 
\begin{equation}
\label{rpm}
 r_\pm=\frac{r_0+r_1}{2} \mp S_z (r_0-r_1),
\end{equation} 
where
\begin{subequations}
\label{r01}
\begin{align}
 r_0{(\omega)} &= -1+\frac{{2}{\rm i}\varkappa_1}{\omega-\omega_c+{\rm i}\varkappa}, \\
  r_1{(\omega)} &=-1+\frac{{2}{\rm i}\varkappa_1}{\omega-\omega_c+{\rm i}\varkappa-\frac{g^2}{\omega-\omega_0+{\rm i}\gamma}}.
\end{align}
\end{subequations}
Here we have introduced $\varkappa_1$ and $\varkappa_2$ being the photon escape rates through the mirrors (light is incident on the mirror characterized by $\varkappa_1$), in these notations {$\varkappa=\varkappa_1+\varkappa_2$}. The transmission coefficients through the microcavity are given by $t_\pm = (1+r_\pm)\sqrt{\varkappa_2/\varkappa_1}$. Equations~\eqref{rpm}, \eqref{r01} can be derived from general Eq.~\eqref{density:m} retaining only states with $0$ and $1$ excitation in the system and neglecting the magnetic field $\Omega$; the magnetic field effect is considered below. These formulae are valid provided that the duration of the probe pulse, $\tau_p$, exceeds by far $\varkappa^{-1}$, so that one can treat it  to be quasi-monochromatic, but, on the other hand, $\tau_p$ should be short enough compared with the electron spin reorientation time, otherwise $S_z$ can not be considered as a constant during the probe pulse action and strong fluctuations of reflectivity and transmission arise~\footnote{The spin noise for a QD in a microcavity operating in a strong coupling regime will be addressed elsewhere.}.

\begin{figure}[t]
\includegraphics[width=0.99\linewidth]{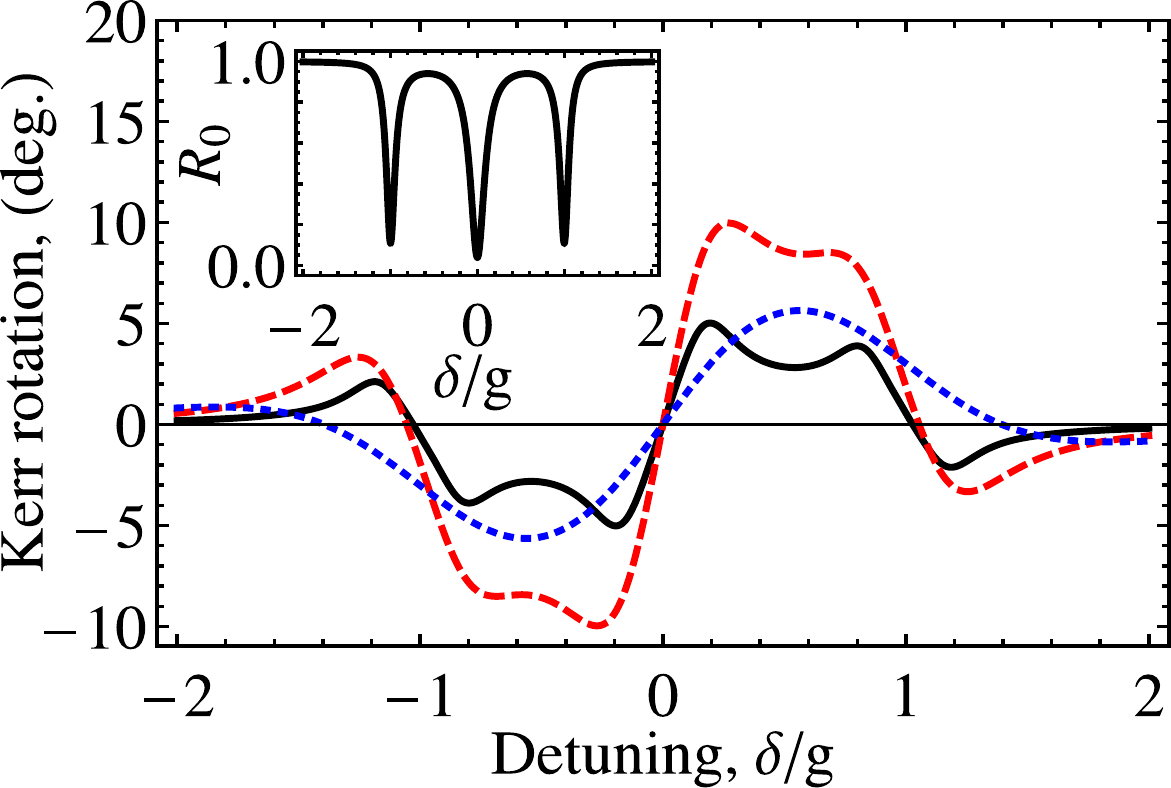}
\caption{Spectra of Kerr angle for completely polarized electron, $S_z^{(0)}=+1/2$.
Parameters of the calculation: $\Omega=g/3$, $\varkappa_1=0.08g$, $\varkappa_2=0.02g$, $\gamma=0.04g$, $g\tau_p=20$ (black solid curves), $10$ (red dashed curves) and $3$ (blue dotted curves). The inset shows the reflectivity spectra in case of unpolarized electron for the continuous monochromatic probe and $\Omega=0$.}
\label{fig:probe}
\end{figure}

For the unpolarized electron the intensity coefficient of light reflection is polarization independent and, according to Eq.~(\ref{rpm}), given by 
\begin{equation}
R_0(\omega) = \frac{1}{4} |r_0(\omega)+r_1(\omega)|^2.
\label{R0}
\end{equation}
Accordingly, the three dips appear in the reflectivity spectrum at a zero-detuning condition $\omega_0=\omega_c$, see inset in Fig.~\ref{fig:probe}. The  dip at the frequency $\omega_c$ corresponds to the light component experiencing no coupling with the QD, its reflectivity is given by $r_0$, and the two dips centered at $\omega\approx\omega_0\pm g$ correspond to the polariton resonances split in the strong coupling regime and described by $r_1$. Noteworthy, this is in contrast with the case of low-density two-dimensional electron gas in the microcavity~\cite{PhysRevB.85.195313}, where for unpolarized electrons two features are present in the reflection/transmission spectra in the strong coupling regime. In the latter case the circularly polarized components of the probe beam always interact with multitudes of electrons.

For the \emph{cw} probe beam the spin-Kerr/Faraday angles and induced ellipticity are given by~\cite{microcavities,glazov:review,yugova09}
\begin{subequations}
\label{KFE}
\begin{equation}
\theta_K(\omega) = \frac{1}{2} \mathrm{Arg}\{r_+(\omega)r_-^*(\omega) \},
\end{equation}
\begin{equation}
 \theta_F(\omega) = \frac{1}{2} \mathrm{Arg}\{t_+(\omega)t_-^*(\omega) \}, 
\end{equation} 
\begin{equation}
\label{ell:r}
 E = \frac{|r_+(\omega)|^2-|r_-(\omega)|^2}{|r_+(\omega)|^2+|r_-(\omega)|^2},
\end{equation}
\end{subequations}
respectively. The ellipticity of the transmitted beam is given by Eq.~\eqref{ell:r} with the replacement $r_\pm \to t_\pm$.

In the case of a short probe pulse the reflectivity coefficient and spin signals should be convoluted with the normalized Fourier transform $\tilde{\mathcal E}(\omega')$ of the probe pulse envelope defined by
\begin{equation}
\tilde{\mathcal E}(\omega') = \mathcal E(\omega')\left[2\pi\int_{-\infty}^\infty \mathrm d t \,  \mathcal E^2(t)\right]^{-1/2},
\end{equation}
where $ \mathcal E(\omega') = \int_{-\infty}^\infty \mathrm d t \, \mathrm e^{\mathrm i (\omega'-\omega) t} \mathcal E(t)$.
The result reads
\begin{subequations}
\label{averaging}
\begin{equation}
{R_0 = \frac{1}{4} \int_{-\infty}^\infty \mathrm d \omega |r_0(\omega)+r_1(\omega)|^2 \tilde{\mathcal E}^2(\omega),}
\end{equation}
\begin{equation}
\label{theta_K}
 \theta_K = \frac{1}{2} \mathrm{Arg}\left\{
 \int_{-\infty}^\infty 
 \mathrm d \omega r_+(\omega)r_-^*(\omega) 
 \tilde{\mathcal E}^2(\omega)\right\},
 \end{equation}
 \begin{equation}
 \theta_F = \frac{1}{2} \mathrm{Arg}\left\lbrace{\int_{-\infty}^\infty \mathrm d \omega t_+(\omega)t_-^*(\omega) \tilde{\mathcal E}^2(\omega)}\right\rbrace,
 \end{equation}
 \begin{equation}
  E = {\frac{\int_{-\infty}^\infty \mathrm d \omega [|r_+(\omega)|^2-|r_-(\omega)|^2]\tilde{\mathcal E}^2(\omega)}{2\int_{-\infty}^\infty \mathrm d \omega [|r_+(\omega)|^2+|r_-(\omega)|^2]\tilde{\mathcal E}^2(\omega)}}.
\end{equation}
\end{subequations}
Note, that the frequency dependences of the Kerr and Faraday rotation signals are similar for the QD embedded into a microcavity operating in a strong coupling regime~\cite{PhysRevB.78.085307,PhysRevA.81.042331}: Both signals consist of the dispersive features centered at polariton and bare cavity mode frequencies and both can reach tens of degrees. The ellipticity has maxima at the eigenfrequencies of the system.

When computing these Kerr and Faraday rotation signals, one can also take into account the effect of magnetic field, $\Omega \gtrsim \varkappa$, similarly to Ref.~\onlinecite{bayer_long}. For a sufficiently weak probe pulse the relevant states involve no more than one photon and are represented by $|\uparrow,0,0\rangle$, $|\downarrow,0,0\rangle$, $|\Uparrow,0,0\rangle$, $|\Downarrow,0,0\rangle$, $|\uparrow,0,1\rangle$, $|\downarrow,1,0\rangle$, $|\uparrow,1,0\rangle$, $|\downarrow,0,1\rangle$. The important feature of the spin coherence detection in this case is that during the lifetime of the probe pulse inside the cavity the electron spin rotates substantially, hence, the reflected and transmitted light is modulated by the spin beats. For the monochromatic probe beam at a frequency $\omega$ the reflected light contains the same-frequency component with the amplitude reduced by the factor 
\begin{equation}
\label{calR0}
\mathcal R_0=r^{(0)}+r^{(x)}S_x,
\end{equation}
and two components at the frequencies $\omega \pm \Omega$ whose amplitudes are proportional to the incident field amplitude with the coefficients
\begin{equation}
\label{calRpm}
\mathcal R_\pm(\sigma)=\sigma {r^{(\pm)}S_\pm^{(0)}},
\end{equation}
respectively, where  $\sigma = \pm1$ for $\sigma^+$/$\sigma^-$ polarized components of the probe beam, $S_\pm^{(0)} = S_z \pm \mathrm i S_y$ and spin components in Eqs.~\eqref{calR0}, \eqref{calRpm} correspond to $t=0$, i.e. the moment of the probe pulse arrival. Here
\begin{subequations}
\label{r:all}
\begin{multline}
 r^{(0)}(\omega)=\frac{1}{4}\left[ r_1(\omega+\Omega/2)+r_0(\omega+\Omega/2)\right.\\
 \left.+r_1(\omega-\Omega/2)+r_0(\omega-\Omega/2) \right],
\end{multline}
\begin{multline}
 r^{(x)}(\omega)=\frac{1}{2}\left[ r_1(\omega+\Omega/2)+r_0(\omega+\Omega/2)\right.\\
 \left.-r_1(\omega-\Omega/2)-r_0(\omega-\Omega/2) \right],
\end{multline}
\begin{equation}
 r^{(\pm)}(\omega)=\frac{1}{2}\left[ r_0(\omega\pm\Omega/2) - r_1(\omega\pm\Omega/2) \right].
\end{equation}
\end{subequations}
Similar equations hold also for the transmitted light. Note that Eqs.~\eqref{calRpm} describes spin-flip Raman scattering of the light in the zero-dimensional microcavity operating in the strong coupling regime~\cite{zapasskii:raman}.

The components of electric field of reflected pulse can be found from Eqs.~\eqref{calR0}, \eqref{calRpm} and \eqref{r:all}. For the pulse linearly polarized along $x$-axis one has
\begin{subequations}
\label{refl:pulse:all}
\begin{equation}
\label{refl:x}
\mathcal E_x'(t) = \frac{1}{2\pi}\int \mathrm d\omega'  \mathcal{E}(\omega') \mathrm e^{-\mathrm i \omega t} \mathcal R_0,
\end{equation}
\begin{multline}
\label{refl:y}
\mathcal E_y'(t) = \frac{\mathrm i}{4\pi}\int \mathrm d\omega'  \mathcal{E}(\omega') \mathrm e^{-\mathrm i \omega t} [\mathcal R_+(+1) \mathrm e^{-\mathrm i \Omega t} \\ + \mathcal R_-(+1) \mathrm e^{+\mathrm i \Omega t} -\mathcal R_+(-1) \mathrm e^{-\mathrm i \Omega t} - \mathcal R_-(-1) \mathrm e^{+\mathrm i \Omega t} ].
\end{multline}
\end{subequations}
Equations~\eqref{refl:pulse:all} together with general definitions of spin signals, Ref.~\cite{glazov:review}, allow us to calculate the spin-Kerr and spin-ellipticity signals.

Figure~\ref{fig:probe} shows the Kerr rotation angle calculated as a function of the detuning for different probe pulse durations. The dispersive features of $\theta_K$ correspond to the frequencies $\omega=\omega_0$ and $\omega_0\pm g$. 

Interestingly the dependence of the rotation angle on the pulse duration at a fixed detuning is nonmonotonic. On one hand, the shorter the pulse, the wider its frequency spectrum, hence, the spectral features become less pronounced with a decrease in the pulse duration. On the other hand, for relatively long pulses $\Omega\tau_p \gtrsim 2\pi$ the spin precession during the pulse action suppresses the signal~\cite{bayer_long}. As a result the optimal sensitivity is reached for the moderate pulse durations $g^{-1}\ll\tau_p\ll\Omega^{-1}$. These results show that giant modifications of the optical polarization, which have recently been experimentally demonstrated for \emph{cw} monochromatic light \cite{Arnold2015}, are possible even in the presence of short non-monochromatic pulses. Furthermore, as envisioned in Ref. \cite{Arnold2015} for the \emph{cw} case, with these short pulses it will also be possible to use generalized polarization measurements in a properly-chosen polarization basis: This will allow approaching the ideal situation where the output polarization states corresponding to spin $|\uparrow \rangle$ and $|\downarrow \rangle$ are orthogonal to each other, and thus distinguishable with a single photon detection event. Such a possibility is at the heart of many theoretical proposals relying on the efficient interaction between a single spin and a single photon for a wide range of applications~\cite{9A,10A,11A,13A,14A,15A,16A,17A,18A,PhysRevB.78.085307} {and can be also interesting in view of generation of photonic states with unconventional statistics~\cite{pulsed}}.

\section{Conclusion}

To conclude, we have presented a theory of single electron spin coherence initialization, detection and manipulation by short optical pulses for a quantum dot/microcavity structure operating in the strong coupling regime. Thereby, the Jaynes-Cummings ladder model is extended to spin-polarized zero-dimensional trion polaritons. It is shown that the spin polarization after a single pump pulse can reach 50\% in a moderate transverse magnetic field. The train of pump pulses is shown to cause the complete spin polarization of the resident electron. The direction of the photoinduced spin depends both on the photon helicity and carrier frequency of the pump pulse. The possibilities for the coherent control of single spin in the strong coupling regime are analyzed. The spin rotation induced by a circular pulse is accompanied, in general, by depolarization, which can be moderate for optimal conditions. The spin-Kerr and Faraday rotation angles detected by short linearly polarized pulses are calculated and shown to reach tens of degrees.

\begin{acknowledgements}
This work was partially supported by the Russian Foundation for Basic Research and DFG in the framework of ICRC TRR160, RF President Grants NSh-1085.2014.2, MD-5726.2015.2, and SP-643.2015.5, RF Government Grant 14.Z50.31.0021 (Leading scientist M.H. Bayer), Dynasty Foundation, St.-Petersburg Government Grant, EU Project SPANGL4Q, and also by the French ANR project SPIQE.
\end{acknowledgements}

\appendix

\section{Resonant approximation for spin pumping}
\label{sec:reson}

The condition $\Omega \gg \varkappa$ in analytical model of spin coherence generation can be relaxed under the assumptions $\delta,\Delta\ll g$, $g\tau_p\gg 1$, and $\varkappa\tau_p\ll1$. In this case the ladder of polariton states is not excited by the pump pulse, hence, $C_m^\pm(t)\equiv0$ for $m\neq0$. On the other hand, the resonant excitation of the bare cavity mode at the first stage leads to the formation of coherent state described by Eq.~\eqref{D:sol}. During the next stage, $t \gg \tau_p$, the phase coherence between excited states is destroyed and the quantum dot in the microcavity can be described, similarly to Ref.~\cite{Poddubny2010}, by the set of kinetic equations for occupancies $N_m$ of photonic states with $m$ photons and spin density matrix for the ground (electron) state. Making use of the general Eq.~\eqref{density:m} one finds that [cf. Ref.~\cite{Zhukov07}]
\begin{subequations}
\label{reson-pump}
\begin{align}
\dot{S}_z(t) &= \Omega S_y(t) + g(t),\\
\dot{S}_y(t) &= {-}\Omega S_z(t), \\
\dot{S}_x(t) &= 0,
\end{align}
\end{subequations}
where $g(t)$ accounts for the spin polarization income from the excited states $\left|\downarrow,m\times\sigma^+\right\rangle$ (we recall that we consider the $\sigma^+$ excitation).
Taking into account that~\cite{Poddubny2010} 
\begin{equation}
\label{Nm}
\dot{N}_m = - 2\varkappa m N_{m} + 2{\varkappa} (m+1) N_{m+1}
\end{equation}
and $g(t)= \varkappa N_{1}$ we obtain from Eq.~\eqref{Lindblad} 
\begin{equation}
 {g(t)=-\varkappa N_{\rm ph}e^{-M}}.
\end{equation} 
Here $N_{\rm ph}(t)=\sum_m m N_m\equiv M\left|\psi_{-1/2}\right|^2$ is the number of photons in the cavity and $M$ is the average number of the excited states in the ladder. It follows from Eq.~\eqref{Nm} that
\begin{equation}
 M=\left|\alpha(\infty)\right|^2 e^{-2{\varkappa} t},
\end{equation} 
where the parameter $\alpha(\infty)$ is defined by Eq.~\eqref{alpha}. Solving the set of Eqs.~\eqref{reson-pump} one obtains instead of Eq.~\eqref{precession0}
\begin{equation}
 S_z(t)=S_z^{(0)}\cos\Omega t+\int\limits_0^t g(t')\cos\Omega(t-t') {\rm d}t'.
 \label{reson}
\end{equation} 
Particularly, in the limiting case of $\Omega=0$ the integral \eqref{reson} can be analytically calculated and
\begin{equation}
 S_z(t)=\frac{1}{4}\left[1-\left(1-4S_z^{(0)}\right) e^{-2{\varkappa} t} \right].
 \label{reson0}
\end{equation} 

We compare in Fig.~\ref{fig:beats}(a) the calculation after Eq.~\eqref{reson0} (green dash-dotted line) with exact numerical solution in the density matrix formalism (black solid line). Similarly the panels (b) and (c) illustrate the agreement between the calculations after Eq.~\eqref{reson} and the two other approaches described above in case of $\Omega>0$. The difference between black and green curves is caused only by the moderate magnitude of the parameter $g\tau_p=5$, where the ladder of polariton states is also weakly excited. We have numerically checked that for the larger value of $g\tau_p=15$ the difference becomes negligible.

\section{Spin pumping by a train of circularly polarized pulses}
\label{sec:train}

In order to address the effect of the pump-pulse train with the repetition period $T_R\gg \tau_p,1/\varkappa$ we follow Refs.~\cite{glazov:review,yugova09,yugova12} and introduce a linear relation between the ``long-living'' electron spin components right before the  pump pulse arrival $\bm S^{(b)}$ and right after the pulse arrival $\bm S^{(a)}$ [note, that in this definition, $\bm S^{(a)} = (S_x^{(0)}, S_y^{(0)}, S_z^{(0)})$ in Eq.~\eqref{precession}]:
\begin{equation}
\label{linear}
\bm S^{(a)} = \hat Q \bm S^{(b)} +  \bm G,
\end{equation}
where the operator $\hat Q$ describes spin transformation and depolarization by the pulse and $\bm G$ describes the spin generation. The matrix elements of $\hat Q$ and the components of $\bm G$ are determined by the pump pulse parameters. The spin precession between the pulses, $0<t<T_R$, is described by the linear transformation $\exp{(-t/\tau_s)}\hat R(\Omega t)$, where the matrix of the operator $\hat R$ can be presented according to Eqs.~\eqref{precession}, \eqref{precession0} in the form:
\begin{equation}
\label{R:prec}
\hat R(x) = 
\begin{pmatrix}
1 & 0 & 0\\
0 & \cos{x} & - \sin{x}\\
0 & \sin{x} & \cos{x}
\end{pmatrix}.
\end{equation}
Correspondingly, the steady-state value of $\bm S^{(a)}$ satisfies the equation
\begin{equation}
\label{train:steady}
\bm S^{(a)} = \exp{(-T_R/\tau_s)}\hat Q \hat R(\Omega T_R) \bm S^{(a)} + \bm G.
\end{equation}
Under the condition of phase synchronization $\Omega T_R = 2\pi N$, where the electron spin makes integer number of revolutions between the pulses, $\hat R(2\pi N)$ reduces to the unit operator and we obtain 
\begin{equation}
\label{train:steady}
\bm S^{(a)} = [1- \exp{(-T_R/\tau_s)}\hat Q]^{-1} \bm G.
\end{equation}

For the sake of illustration we consider the resonant case where the pump carrier frequency equals to that of the bare cavity mode, $\omega={\omega_c=}\omega_0$. Then under conditions $g\tau_p\gg1$ and \eqref{b:field} the operator $\hat Q$ and vector $\bm g$ are expressed only via one parameter $0\leqslant Q \leqslant 1$ as [cf. Ref.~\cite{yugova09}]
\begin{equation}
\label{Q:op}
\hat Q = \begin{pmatrix}
Q &0 &0\\
0 & Q& 0\\
0 & 0 & \frac{1}{2}(Q^2+1)
\end{pmatrix},
\end{equation}
and $G=(0,0,(1-Q^2)/4)$. In this case $S_x^{(a)}=S_x^{(b)} = S_y^{(a)}=S_y^{(b)}=0$ and 
\begin{equation}
S_z^{(b)} = \frac{1}{2} \frac{1-Q^2}{2\exp{(T_R/\tau_s)}- 1- Q^2},
\end{equation}
\[
S_z^{(a)} = \frac{1}{2} \frac{1-Q^2}{2-(Q^2+1)\exp{(T_R/\tau_s)}}.
\]
For $T_R \ll \tau_s$, one has $S_z^{(b)} = S_z^{(1)} = 1/2$ which means that the complete spin polarization by the train of pump pulses is achievable.

\section{Rabi cycle in the very strong coupling limit}
\label{sec:Rabi}

In the limit of very strong coupling where $g\gg \varkappa,\gamma$ and for very spectrally narrow circularly polarized pulse $g\tau_p \gg 1$ one can, neglecting magnetic field effect, expect Rabi cycle resulting in the oscillations between the ground state $\uparrow$ and one of the polariton states, see Eq.~\eqref{polaritons} and Fig.~\ref{fig:scheme:tr}, 
\begin{align}
|1,+\rangle = \frac{1}{\sqrt{2}} \left[ |\Uparrow\rangle + |\uparrow, 1\times \sigma^+\rangle\right],\\
|1,-\rangle = \frac{1}{\sqrt{2}} \left[ |\Uparrow\rangle - |\uparrow, 1\times \sigma^+\rangle\right],
\end{align}
if pump pulse is almost resonant with the corresponding transition frequency $\omega_\pm = \omega_0 \pm g$, 
\begin{equation}
\label{detun:Rabi}
|\omega-\omega_+|~~\mbox{or}~~|\omega-\omega_-|\ll g,
\end{equation}
Here, for simplicity, we assumed that $\omega_0 = \omega_c$. In this situation the wavefunction of the system can be presented in the form
\begin{equation}
\label{wave:Rabi}
\Psi = \mathcal A(t) |\uparrow\rangle + \mathcal B(t) |1,\pm\rangle,
\end{equation}
where coefficients $\mathcal A(t)$ and $\mathcal B(t)$ satisfy standard equations for the driven two-level system dynamics~\cite{ll3_eng}:
\begin{align}
\label{Rabi:2level}
\mathrm i \dot{\mathcal A} &= \frac{\mathcal E_+(t)}{\sqrt{2}} \exp{[\mathrm i (\omega - \omega_{\pm})]} \mathcal B,\\
\mathrm i \dot{\mathcal B} &= \omega_\pm \mathcal B + \frac{\mathcal E_+(t)}{\sqrt{2}}\exp{[-\mathrm i (\omega - \omega_{\pm})]} \mathcal A,
\end{align}
and time argument is omitted for brevity. Unlike Eqs.~\eqref{C} of the main text, here, we use here the basis of polariton eigenstates~\eqref{spectrum}. The possibility to disregard other states in Eq.~\eqref{wave:Rabi} is justified for moderate pump pulse areas, $\Theta$, by the condition \eqref{detun:Rabi}: The transition from either of $|1,\pm\rangle$ states to the higher energy states are suppressed since corresponding transition frequencies differ strongly from $\omega_\pm$. An increase of the pump pulse area $\Theta \gg 2\pi$, where in two-level approximation many periods of Rabi cycle would take place, invokes here the transitions to the higher energy states and results in the damping of Rabi oscillations. We also note, that for the parameters of calculations presented in Figs.~\ref{fig:pump},~\ref{fig:rot} many excited states are involved and Rabi oscillations do not take place.

Inclusion of the magnetic field induced mixing of the states $|\uparrow\rangle$, $|\downarrow\rangle$ results in more complex picture of Rabi oscillations, cf. Ref.~\cite{bayer_long}.

It is also worths mentionning that control pulses of specific shapes, e.g., rectangular pulses, under certain conditions, namely, $0<\left|\delta\right|\ll g$ and  $\delta\tau_p\sim 1$, can lead to the spin rotation without any depolarization. Indeed, the parameter $\alpha$ in Eq.~\eqref{D:sol} for $D_m$ can be made zero for specific pulse duration while the phase $\phi$ is not, generally, a multiple of $2\pi$. In this cases $D_0$ remains constant in amplitude but acquires the phase, while $C_0^-$ is unaffected by the control pulse.

We note that in the limit of weak coupling where the feedback of the quantum dot on the cavity mode can be neglected (and the field inside the cavity can be treated classically), Rabi oscillations naturally arise since photons of given circular, $\sigma^+$ ($\sigma^-$) polarizations couple $|\uparrow\rangle$ ($|\downarrow\rangle$) electron with $|\Uparrow\rangle$ ($\Downarrow\rangle$) trion states, see Refs.~\cite{yugova09,glazov:review} for more details.


\begin{thebibliography}{100}
\providecommand{\selectlanguage}[1]{\relax}
\newcommand{\bbletal}[1]{et al}
\newcommand{\Capitalize}[1]{\uppercase{#1}}
\newcommand{\capitalize}[1]{\expandafter\Capitalize#1}

\bibitem{dyakonov_book}
M.~I. Dyakonov (ed.).
\newblock \emph{Spin physics in semiconductors} (Springer-Verlag: Berlin,
  Heidelberg, 2008).

\bibitem{PhysRevLett.94.047402}
A.~S. Bracker, E.~A. Stinaff, D.~Gammon, M.~E. Ware, J.~G. Tischler,
  A.~Shabaev, A.~L. Efros, D.~Park, D.~Gershoni, V.~L. Korenev, I.~A. Merkulov.
\newblock \emph{Optical Pumping of the Electronic and Nuclear Spin of Single
  Charge-Tunable Quantum Dots}.
\newblock Phys. Rev. Lett. \textbf{94}, 047402 (2005).

\bibitem{greilich06}
A.~Greilich, R.~Oulton, E.~A. Zhukov, I.~A. Yugova, D.~R. Yakovlev, M.~Bayer,
  A.~Shabaev, A.~L. Efros, I.~A. Merkulov, V.~Stavarache, D.~Reuter, A.~Wieck.
\newblock \emph{Optical Control of Spin Coherence in Singly Charged
  $\mbox{(In,Ga)As/GaAs}$ Quantum Dots}.
\newblock Phys. Rev. Lett. \textbf{96}, 227401 (2006).

\bibitem{Greilich2009}
A.~Greilich, S.~E. Economou, S.~Spatzek, D.~R. Yakovlev, D.~Reuter, A.~D.
  Wieck, T.~L. Reinecke, M.~Bayer.
\newblock \emph{Ultrafast optical rotations of electron spins in quantum dots}.
\newblock Nature Physics \textbf{5}, 262 (2009).

\bibitem{A.Greilich07212006}
A.~Greilich, D.~R. Yakovlev, A.~Shabaev, A.~L. Efros, I.~A. Yugova, R.~Oulton,
  V.~Stavarache, D.~Reuter, A.~Wieck, M.~Bayer.
\newblock \emph{{Mode locking of electron spin coherences in singly charged
  quantum dots}}.
\newblock Science \textbf{313}, 341 (2006).

\bibitem{A.Greilich09282007}
A.~Greilich, A.~Shabaev, D.~R. Yakovlev, A.~L. Efros, I.~A. Yugova, D.~Reuter,
  A.~D. Wieck, M.~Bayer.
\newblock \emph{{Nuclei-induced frequency focusing of electron spin
  coherence}}.
\newblock Science \textbf{317}, 1896 (2007).

\bibitem{glazov:review}
M.~M. Glazov.
\newblock \emph{Coherent spin dynamics of electrons and excitons in
  nanostructures (a review)}.
\newblock Physics of the Solid State \textbf{54}, 1 (2012).

\bibitem{MeteAtature04282006}
M.~Atature, J.~Dreiser, A.~Badolato, A.~Hogele, K.~Karrai, A.~Imamoglu.
\newblock \emph{{Quantum-Dot Spin-State Preparation with Near-Unity Fidelity}}.
\newblock Science \textbf{312}, 551 (2006).

\bibitem{J.Berezovsky04182008}
J.~Berezovsky, M.~H. Mikkelsen, N.~G. Stoltz, L.~A. Coldren, D.~D. Awschalom.
\newblock \emph{{Picosecond Coherent Optical Manipulation of a Single Electron
  Spin in a Quantum Dot}}.
\newblock Science \textbf{320}, 349 (2008).

\bibitem{mikkelsen07}
M.~H. Mikkelsen, J.~Berezovsky, N.~G. Stoltz, L.~A. Coldren, D.~D. Awschalom.
\newblock \emph{Optically detected coherent spin dynamics of a single electron
  in a quantum dot}.
\newblock Nature Physics \textbf{3}, 770 (2007).

\bibitem{atature07}
M.~Atature, J.~Dreiser, A.~Badolato, A.~Imamoglu.
\newblock \emph{Observation of $\mbox{F}$araday rotation from a single confined
  spin}.
\newblock Nature Physics \textbf{3}, 101 (2007).

\bibitem{2011NaPho...5..702G}
A.~{Greilich}, S.~G. {Carter}, D.~{Kim}, A.~S. {Bracker}, D.~{Gammon}.
\newblock \emph{{Optical control of one and two hole spins in interacting
  quantum dots}}.
\newblock Nature Photonics \textbf{5}, 702 (2011).

\bibitem{Kavokin:1997fk}
A.~V. Kavokin, M.~R. Vladimirova, M.~A. Kaliteevski, O.~Lyngnes, J.~D. Berger,
  H.~M. Gibbs, G.~Khitrova.
\newblock \emph{Resonant Faraday rotation in a semiconductor microcavity}.
\newblock Phys. Rev. B \textbf{56}, 1087 (1997).

\bibitem{PhysRevB.78.085307}
C.~Y. Hu, A.~Young, J.~L. O'Brien, W.~J. Munro, J.~G. Rarity.
\newblock \emph{Giant optical Faraday rotation induced by a single-electron
  spin in a quantum dot: Applications to entangling remote spins via a single
  photon}.
\newblock Phys. Rev. B \textbf{78}, 085307 (2008).

\bibitem{PhysRevB.85.195313}
R.~Giri, S.~Cronenberger, M.~Vladimirova, D.~Scalbert, K.~V. Kavokin, M.~M.
  Glazov, M.~Nawrocki, A.~Lema\^{i}tre, J.~Bloch.
\newblock \emph{Giant photoinduced Faraday rotation due to the spin-polarized
  electron gas in an $n$-GaAs microcavity}.
\newblock Phys. Rev. B \textbf{85}, 195313 (2012).

\bibitem{PhysRevLett.111.087603}
R.~Giri, S.~Cronenberger, M.~M. Glazov, K.~V. Kavokin, A.~Lema\^{i}tre,
  J.~Bloch, M.~Vladimirova, D.~Scalbert.
\newblock \emph{Nondestructive Measurement of Nuclear Magnetization by
  Off-Resonant Faraday Rotation}.
\newblock Phys. Rev. Lett. \textbf{111}, 087603 (2013).

\bibitem{De-Greve:2011uq}
K.~De~Greve, P.~L. McMahon, D.~Press, T.~D. Ladd, D.~Bisping, C.~Schneider,
  M.~Kamp, L.~Worschech, S.~Hofling, A.~Forchel, Y.~Yamamoto.
\newblock \emph{Ultrafast coherent control and suppressed nuclear feedback of a
  single quantum dot hole qubit}.
\newblock Nat Phys \textbf{7}, 872 (2011).

\bibitem{PhysRevLett.112.156601}
R.~Dahbashi, J.~H\"ubner, F.~Berski, K.~Pierz, M.~Oestreich.
\newblock \emph{Optical Spin Noise of a Single Hole Spin Localized in an
  (InGa)As Quantum Dot}.
\newblock Phys. Rev. Lett. \textbf{112}, 156601 (2014).

\bibitem{Arnold2015}
C.~Arnold, J.~Demory, V.~Loo, A.~Lema\^{i}tre, I.~Sagnes, M.~Glazov, O.~Krebs,
  P.~Voisin, P.~Senellart, L.~Lanco.
\newblock \emph{Macroscopic rotation of photon polarization induced by a single
  spin}.
\newblock Nat Commun \textbf{6} (2015).

\bibitem{andreani99a}
L.~C. Andreani, G.~Panzarini, J.-M. G\'erard.
\newblock \emph{Strong-coupling regime for quantum boxes in pillar
  microcavities: Theory}.
\newblock Phys. Rev. B \textbf{60}, 13276 (1999).

\bibitem{Khitrova2006}
G.~Khitrova, H.~M. Gibbs, M.~Kira, S.~W. Koch, A.~Scherer.
\newblock \emph{Vacuum Rabi splitting in semiconductors}.
\newblock Nat Phys \textbf{2}, 81 (2006).

\bibitem{microcavities}
A.~Kavokin, J.~Baumberg, G.~Malpuech, F.~Laussy.
\newblock \emph{Microcavities} (Oxford University Press, UK, 2011).

\bibitem{ivchenko05a}
E.~L. Ivchenko.
\newblock \emph{Optical spectroscopy of semiconductor nanostructures} (Alpha
  Science, Harrow UK, 2005).

\bibitem{yoshie04a}
T.~Yoshie, A.~Scherer, J.~Heindrickson, G.~Khitrova, H.~M. Gibbs, G.~Rupper,
  C.~Ell, O.~B. Shchekin, D.~G. Deppe.
\newblock \emph{Vacuum {Rabi} splitting with a single quantum dot in a photonic
  crystal nanocavity}.
\newblock Nature \textbf{432}, 200 (2004).

\bibitem{reithmaier04a}
J.~P. Reithmaier, G.~Sek, A.~L\"offler, C.~Hofmann, S.~Kuhn, S.~Reitzenstein,
  L.~V. Keldysh, V.~D. Kulakovskii, T.~L. Reinecker, A.~Forchel.
\newblock \emph{Strong coupling in a single quantum dot--semiconductor
  microcavity system}.
\newblock Nature \textbf{432}, 197 (2004).

\bibitem{peter05a}
E.~Peter, P.~Senellart, D.~Martrou, A.~Lematre, J.~Hours, J.~M. Gerard,
  J.~Bloch.
\newblock \emph{Exciton-Photon Strong-Coupling Regime for a Single Quantum Dot
  Embedded in a Microcavity}.
\newblock Phys. Rev. Lett. \textbf{95}, 67401 (2005).

\bibitem{hennessy07}
K.~Hennessy, A.~Badolato, M.~Winger, D.~Gerace, M.~Atat\"{u}re, S.~Gulde,
  S.~F\"{a}lt, E.~L. Hu, A.~Imamo\u{g}lu.
\newblock \emph{Quantum nature of a strongly coupled single quantum dot-cavity
  system}.
\newblock Nature \textbf{445}, 896 (2007).

\bibitem{JC} E.T. Jaynes, F.W. Cummings. \emph{Comparison of quantum and semiclassical radiation theories with application to the beam maser.} Proc. IEEE {\bf 51}, 89 (1963). 

\bibitem{atomcavity} I. Schuster, A. Kubanek, A. Fuhrmanek, T. Puppe, P. W. H. Pinske, K. Murr, and G. Rempe, 
\newblock \emph{Nonlinear spectroscopy of photons bound
to one atom}. \newblock Nature Phys. {\bf 4}, 382 (2008).

\bibitem{fink} J. M. Fink, M. G\"{o}ppl, M. Baur, R. Bianchetti, P. J. Leek, A. Blais, and A. Wallraff. \emph{Climbing the Jaynes-Cummings ladder and observing its $\sqrt{n}$ nonlinearity in a cavity QED system}. Nature {\bf 454}, 315 (2008).

\bibitem{dotcavity} J. Kasprzak, S. Reitzenstein, E. A. Muljarov, C. Kistner, C. Schneider, M. Strauss, S. H\"ofling,
A. Forchel, and W. Langbein, Nature Mater. {\bf 9}, 304 (2010).
\newblock \emph{Up on the Jaynes√Cummings ladder of a quantum-dot/microcavity system}. \newblock Nature Phys. {\bf 4}, 382 (2008).

\bibitem{Mar99}
X.~Marie, T.~Amand, P.~Le~Jeune, M.~Paillard, P.~Renucci, L.~E. Golub, V.~D.
  Dymnikov, E.~L. Ivchenko.
\newblock \emph{Hole spin quantum beats in quantum-well structures}.
\newblock Phys. Rev. B \textbf{60}, 5811 (1999).

\bibitem{Poddubny2010}
A.~N. Poddubny, M.~M. Glazov, N.~S. Averkiev.
\newblock \emph{Nonlinear emission spectra of quantum dots strongly coupled to
  a photonic mode}.
\newblock Phys. Rev. B \textbf{82}, 205330 (2010).

\bibitem{milburn}
D.~F. Walls, G.~J. Milburn.
\newblock \emph{Quantum optics} (Springer, 2008), 2nd ed.

\bibitem{Carmichael}
H.~Carmichael.
\newblock \emph{An open system approach to quantum optics} (Springer-Verlag,
  1993).
  
\bibitem{deph2} {A. Laucht, N. Hauke, J. M. Villas-B\^{o}as, F. Hofbauer, G. B\"{o}hm, M. Kaniber, and J. J. Finley, \emph{Dephasing of Exciton Polaritons in Photoexcited InGaAs Quantum Dots in GaAs Nanocavities}, Phys. Rev. Lett. {\bf 103}, 087405 (2009).}  
  
\bibitem{deph1} {B. Krummheuer, V. M. Axt, and T. Kuhn, \emph{Theory of pure dephasing and the resulting absorption line shape in semiconductor quantum dots}, Phys. Rev. B {\bf 65}, 195313 (2002).}



\bibitem{deph3} {A. Auff\`{e}ves, D. Gerace, J.-M. G\'{e}rard, M. Franca Santos, L. C. Andreani, and J.-P. Poizat, \emph{Controlling the dynamics of a coupled atom-cavity system by pure dephasing}, Phys. Rev. {\bf B 81}, 245419 (2010).}

  

\bibitem{pulsed}
{N.~H. Lindner, T. Rudolph.
\newblock \emph{Proposal for Pulsed On-Demand Sources of Photonic Cluster State Strings}.
\newblock Phys. Rev. Lett. \textbf{103}, 113602 (2009).}

\bibitem{non-markovian}
{D.~P.~S. McCutcheon, N.~H. Lindner, T. Rudolph.
\newblock \emph{Error Distributions on Large Entangled States with Non-Markovian Dynamics}.
\newblock Phys. Rev. Lett. \textbf{113}, 260503 (2014).}

\bibitem{just} {E. del Valle, F. P. Laussy, and C. Tejedor, \emph{Luminescence spectra of quantum dots in microcavities. II. Fermions}, Phys. Rev. B {\bf 79}, 235326 (2009); N. S. Averkiev, M. M. Glazov, and A. N. Poddubnyi, \emph{Collective Modes of Quantum Dot Ensembles in Microcavities}, JETP {\bf 108}, 836 (2009).}

\bibitem{Laussy} F. P. Laussy, E. del Valle, M. Schrapp, A. Laucht, and J. J. Finley. \emph{Climbing the Jaynes-Cummings ladder by photon counting}. Journal of Nanophotonics, {\bf 6}, 061803 (2012).



\bibitem{yugova09}
I.~A. Yugova, M.~M. Glazov, E.~L. Ivchenko, A.~L. Efros.
\newblock \emph{Pump-probe Faraday rotation and ellipticity in an ensemble of
  singly charged quantum dots}.
\newblock Phys. Rev. B \textbf{80}, 104436 (2009).

\bibitem{glauber63}
R.~J. Glauber.
\newblock \emph{Coherent and Incoherent States of Radiation Field}.
\newblock Phys. Rev. \textbf{131}, 2766 (1963).

\bibitem{poshakinskii}
A.V. Poshakinskiy and A. N. Poddubny. \emph{Time-dependent photon correlations for incoherently pumped quantum dot strongly coupled to the cavity mode}. JETP {\bf 145}, 237 (2014).

\bibitem{PhysRev.40.502}
N.~Rosen, C.~Zener.
\newblock \emph{Double $\mbox{Stern-Gerlach}$ Experiment and Related Collision
  Phenomena}.
\newblock Phys. Rev. \textbf{40}, 502 (1932).

\bibitem{ll3_eng}
L.~Landau, E.~Lifshitz.
\newblock {\selectlanguage{english}\emph{Quantum Mechanics: Non-Relativistic
  Theory (vol. 3)}} (Butterworth-Heinemann, Oxford, 1977).

\bibitem{yugova12}
I.~A. Yugova, M.~M. Glazov, D.~R. Yakovlev, A.~A. Sokolova, M.~Bayer.
\newblock \emph{Coherent spin dynamics of electrons and holes in semiconductor
  quantum wells and quantum dots under periodical optical excitation: Resonant
  spin amplification versus spin mode locking}.
\newblock Phys. Rev. B \textbf{85}, 125304 (2012).

\bibitem{bayer_long}
S.~Spatzek, S.~Varwig, M.~M. Glazov, I.~A. Yugova, A.~Schwan, D.~R. Yakovlev,
  D.~Reuter, A.~D. Wieck, M.~Bayer.
\newblock \emph{Generation and detection of mode-locked spin coherence in
  $\mbox{(In,Ga)As/GaAs}$ quantum dots by laser pulses of long duration}.
\newblock Phys. Rev. B \textbf{84}, 115309 (2011).

\bibitem{carter:167403}
S.~G. Carter, A.~Shabaev, S.~E. Economou, T.~A. Kennedy, A.~S. Bracker, T.~L.
  Reinecke.
\newblock \emph{Directing Nuclear Spin Flips in $\mbox{InAs}$ Quantum Dots
  Using Detuned Optical Pulse Trains}.
\newblock Phys. Rev. Lett. \textbf{102}, 167403 (2009).

\bibitem{spherical-dots}
D.~S. Smirnov, M.~M. Glazov.
\newblock \emph{Spin coherence generation and detection in spherical
  nanocrystals}.
\newblock Journal of Physics: Condensed Matter \textbf{24}, 345302 (2012).

\bibitem{PhysRevA.81.042331}
N.~Takei, M.~Takeuchi, Y.~Eto, A.~Noguchi, P.~Zhang, M.~Ueda, M.~Kozuma.
\newblock \emph{Faraday rotation with a single-nuclear-spin qubit in a
  high-finesse optical cavity}.
\newblock Phys. Rev. A \textbf{81}, 042331 (2010).

\bibitem{Zhukov07}
E.~A. Zhukov, D.~R. Yakovlev, M.~Bayer, M.~M. Glazov, E.~L. Ivchenko,
  G.~Karczewski, T.~Wojtowicz, J.~Kossut.
\newblock \emph{Spin coherence of a two-dimensional electron gas induced by
  resonant excitation of trions and excitons in CdTe/(Cd,Mg)Te quantum wells}.
\newblock Phys. Rev. B \textbf{76}, 205310 (2007).

\bibitem{zapasskii:raman} M. M. Glazov and V. S. Zapasskii. \emph{Linear optics, Raman scattering, and spin noise spectroscopy}. Optics Express {\bf 23}, 11713 (2015).

\bibitem{9A} C.Y.~Hu, W.J.~Munro, and J.G.~Rarity. \emph{Deterministic photon entangler using a charged quantum dot inside a microcavity}. Phys. Rev. B {\bf 78}, 125318 (2008). 
\bibitem{10A} C.~Bonato, et al. \emph{CNOT and Bell-state analysis in the weak-coupling cavity QED regime}. Phys. Rev. Lett. {\bf 104}, 160503 (2010).
\bibitem{11A} M.N.~Leuenberger. \emph{Fault-tolerant quantum computing with coded spins using the conditional Faraday rotation in quantum dots.} Phys. Rev. B {\bf 73}, 075312 (2006).
\bibitem{13A} Y. Takahashi, et al. \emph{Quantum nondemolition measurement of spin via the paramagnetic Faraday rotation}. Phys. Rev. A {\bf 60}, 4974√4979 (1999). 
\bibitem{14A} M.N.~Leuenberger, M.E. Flatt\'{e}, D.D. Awschalom. \emph{Teleportation of electronic many-qubit states encoded in the electron spin of quantum dots via single photons}. Phys. Rev. Lett. {\bf 94}, 107401 (2005).
\bibitem{15A} E. Waks and J. Vuckovic. \emph{Dipole induced transparency in drop-filter cavity- waveguide systems}. Phys. Rev. Lett. {\bf 96}, 153601 (2006).
\bibitem{16A} S. K. Y.~Lee and C. K. Law. \emph{Analysis of photon-atom entanglement generated by Faraday rotation in a cavity.} Phys. Rev. A {\bf 73}, 053808 (2006).
\bibitem{17A} C.Y.~Hu, J.G.~Rarity. \emph{Loss-resistant state teleportation and entanglement swapping using a quantum-dot spin in an optical microcavity.} Phys. Rev. B {\bf 83}, 115303 (2011).
\bibitem{18A} A. B.~Young, C.Y.~Hu, J.G.~ Rarity. \emph{Generating entanglement with low-Q-factor microcavities.} Phys. Rev. A {\bf 87}, 012332 (2013).

\end{thebibliography}
\end{document}